**Mechanisms of de-icing by surface Rayleigh and plate Lamb acoustic waves**

*Shilpi Pandey,\* Jaime del Moral,\* Stefan Jacob, Laura Montes, Jorge Gil-Rostra, Alejandro Frechilla, Atefeh Karimzadeh, Victor J. Rico, Raul Kantar, Niklas Kandelin, Carmen López-Santos, Heli Koivuluoto, Luis Angurel, Andreas Winkler,\* Ana Borrás, Agustin R. González-Elipe.*

S. Pandey,\*[+] S. Jacob, A. Karimzadeh, A. Winkler\*

Leibniz IFW DresdenSAWLab Saxony,

RG Acoustic MicrosystemsHelmholtzstr. 20, 01069 Dresden, Germany

\*E-mail: (shilpi.pandey@tum.de) (a.winkler@ifw-dresden.de)

S.Jacob

German National Metrology Institute (PTB), Bundesallee 100, 38106 Braunschweig, Germany

J. D. Moral,\* L. Montes, J. G. Rostra, V. J. Rico, C. L. Santos, A. Borrás, A. R. González-Elipe.

Nanotechnology on Surfaces and Plasma Lab, Materials Science Institute of Seville, Consejo Superior de Investigaciones Científicas (CSIC), Americo Vespucio 49, Sevilla 41092, Spain

\*E-mail: (jaime.delmoral@icmse.csic.es)

R. Kantar, N. Kandelin, H. Koivuluoto

Materials Science and Environmental Engineering, Faculty of Engineering and Natural Sciences, Tampere University, 589, 33104 Tampere, Finland

A. Frechilla, L. A. Angurel

Instituto de Nanociencia y Materiales de Aragón (INMA), CSIC-Universidad de Zaragoza, Zaragoza, Spain

*Current addresses*:

[+] Heinz-Nixdorf-Chair of Biomedical Electronics,  School of Computation, Information and Technology. Technical University of Munich, TranslaTUM, 80333 Munich, Germany

Acoustic waves (AW) have recently emerged as an energy-efficient ice removal procedure compatible with functional and industrial-relevant substrates. However, critical aspects at fundamental and experimental levels have yet to be disclosed to optimize their operational conditions. Identifying the processes and mechanisms by which different types of AWs induce de-icing are some of these issues. Herein, using model $LiNbO_3$ systems and two types of interdigitated transducers, we analyze the de-icing and anti-icing efficiencies and mechanisms driven by Rayleigh surface acoustic waves (R-SAW) and Lamb waves with 120 and 510 µm wavelengths, respectively. Through the experimental analysis of de-icing and active anti-icing processes and the finite element simulation of the AW generation, propagation, and interaction with small ice aggregates, we disclose that Lamb waves are more favorable than R-SAWs to induce de-icing and/or prevent the freezing of droplets. Prospects for applications of this study are supported by proof of concept experiments, including de-icing in an ice wind tunnel, demonstrating that Lamb waves can efficiently remove ice layers covering large LN substrates. Results indicate that the de-icing mechanism may differ for Lamb waves or R-SAWs and that the wavelength must be considered as an important parameter for controlling the efficiency.

### 3.3.  1. Introduction

Ice accretion on materials operating outdoors or in sub-zero environments is a critical issue that significantly impacts efficiency, maintenance, and security in various industries, including aviation, energy generation, and the functionality of sensor and camera windows and screens. To address this problem, two primary approaches have been proposed: i) *Passive anti-icing* based on applying surface engineering processes to reduce or delay ice accretion;[1-5] very often, these approaches rely on superhydrophobic or liquid-infused surfaces.[6-9] ii) *Active de-icing/anti-icing* systems applying different strategies for ice removal or preventing its formation; recent advances in the field include nanotechnology-based melting by Joule heating,[10] using materials such as graphene,[11-13] or photothermal de-icing using plasmonic or magnetic nanoparticles.[14-18] An emergent de-icing approach involves exciting surfaces by high-frequency acoustic waves with nano-scale amplitudes, a procedure compatible with industrial-relevant surfaces. In this context, a key distinction should be made between the localized application to surfaces of commercially available ultrasound generators[19-20] and the innovative concepts based on the incorporation of piezoelectric active supports or layers, as well as electrodes (lateral field excitation (LFE) or interdigitated transducers (IDTs)), as integral parts of the surface to be activated.[21-26] The present work fits within this second approach, dealing with the surface integration of surface or bulk acoustic waves (SAWs and BAWs) for de-icing and ice sensing. Among the articles addressing this topic, the work of Yang et al. (2020), which proposes a strategy to weaken ice adhesion using Rayleigh surface acoustic waves (R-SAW),[21] is particularly noteworthy. For this purpose, IDTs with different wavelengths (100, 200, 300, and 400 µm) were manufactured on a 5 µm thick ZnO piezoelectric thin film deposited on an aluminum substrate, producing SAWs with frequencies comprised between 7.38 and 27.84 MHz. Authors state that SAWs produce vibrations comparable to "nano-scale earthquakes", which result in the generation of acoustic-heating effects, the development of micro-cracks, and the disturbance of ice nucleation through local vibrations and energy streaming into the formed liquid phase. The contribution of Joule heating effects directly stemming from ohmic losses in the IDTs cannot be discarded in these experiments since ice droplets were placed directly atop the IDTs. The same authors have recently claimed the possibility of monitoring ice formation using this type of SAWs.[22] Similarly, Zeng et al. 2021,[23] demonstrated a decrease in adhesion between ice and substrate due to interface heating and a reduction in electrostatic forces and mechanical interlocking on SAW-activated substrates. Del Moral et al.,[24] and Jacob et al.,[25] have recently extended the application of acoustic wave de-icing to transparent substrates and supports. In the first work, authors demonstrated that extended electrodes on piezoelectric plates in an LFE configuration generate Thickness Shear Mode - Bulk Acoustic Waves (TSM - BAWs) in the 3-4 MHz range. These waves efficiently induced de-icing and a reduction of ice accretion (i.e., active-anti-icing effect), as well as a decrease in ice adhesion. A drawback of this approach is the necessity of a fast-electronic tuning of the excitation due to the very narrow bandwidth of BAW modes. The possibility of using the same device for ice detection under realistic conditions of operation (i.e., experiments in icing wind tunnels (IWT)) and a profitable synergy to reduce power consumption upon the application of anti-icing coatings were also demonstrated in Del Moral et al.[24] On the other hand, Jacob et al.[25] demonstrated that R-SAW technology can be applied to large areas and transparent substrates beyond the centimeter scale, proving that R-SAWs can be used for de-icing industrially relevant surfaces. In that work, the authors also demonstrated that R-SAWs generated on piezoelectric plates and piezoelectric thin films induced the de-icing of glaze ice through a pure acoustic mechanism, discarding the melting through direct electrothermal (Joule) heating induced by ohmic losses in the IDTs on substrates with low thermal conductivity. Recently, the same authors demonstrated that a hybrid operational mode is possible by combining R-SAW for de-icing and BAW for sensing by carefully controlling the excitation mode of the two finger combs integrated into the IDTs.[26] Other relevant studies on the

implementation of AW ice sensors and de-icing of rime ice have also been recently published.[27-30]

Despite the significance of these contributions, de-icing with substrate-integrated AW systems faces substantial fundamental uncertainties that still hinder the implementation of this technology in real-world applications. Critical bottlenecks encompass the assessment of factors affecting the transmission of AW energy to ice or water present on the device surface, understanding the mechanisms behind ice-cracking and other processes contributing to ice removal without complete melting, and defining activation conditions to avoid ice formation under subzero temperatures, thereby enabling an efficient anti-icing mode. Additional features preventing an unequivocal comparison of efficiencies and de-icing mechanisms are the differences in the type of AWs, either SAWs, Lamb waves or TSM-BAWs,[21-26] the effect of the substrate materials (e.g., transparent and black LN, ZnO thin films deposited on different substrates such as Al foils, glass and fused silica, the hydrophobic or hydrophilic surface modification of these substrates, etc.), or the plethora of conditions used in the experiments (type of ice, liquid water content (LWC) when working in IWTs, static and windy conditions, temperatures and relative humidity, water freezing or ice accretion, etc.).

Herein, trying to shed some light on this twilight scenario, we systematically investigate the de-icing and active anti-icing activation of the same substrate with either R- SAW or Lamb waves. For this purpose, we use $LiNbO_3$ (LN) piezoelectric chips activated with IDTs operating at different driving frequencies. The selection of LN for preparing specific model systems to study de-icing with AWs exceeds this topic, as this piezoelectric material is widely utilized in other applications such as high-temperature sensors or optical waveguide tapered antennas.[31-33] Herein, we have electro-acoustically characterized a series of IDT configurations to optimize their performance, either for activating R-SAWs or Lamb waves. Then, for these two wave modes and various sub-zero temperatures, we performed de-icing and active anti-icing experiments of small ice aggregates and water droplets, monitoring the threshold power required in each case and on a larger amount of ice generated in an ice wind tunnel. Additionally, an analysis of the operational conditions of the chips using finite element model (FEM) simulations in COMSOL Multiphysics is used to model the generation of the R-SAW and Lamb wave modes and account for their interaction with ice. Finally, to validate these experimental and simulation results for Lamb waves, we have carried out some proof of concept de-icing experiments proving the suitability of the Lamb wave activation to remove large areas of accreted ice. The overall assessment of the analysis of the de-icing of small ice aggregates and large ice layers with Lamb waves has permitted the proposal of a specific mechanism of de-icing that differs from that already known for R-SAWs.[25, 30] We are confident that all these findings will significantly contribute to establishing a rational pathway for exploiting AWs as the basis for a next-generation de-icing system.

**3.4.  2. Materials and Methods**

**3.5.  2.1. Fabrication and electroacoustic characterization of AW chip devices**

The selected piezoelectric material was a 0.5 mm thick black Lithium Niobate 128º YX cut purchased from CSIMC-Freqcontrol, China. The 4-inch wafers were cleaned with acetone and IPA using ultrasonic cleaning for 5 mins and $O_2$ plasma treatment at 100 W for 10 mins. IDTs were patterned using photoresist spin-coating, a maskless laser writer (Heidelberg instruments MLA 100), and the lift-off technique. Ti (5 nm)/Al (295 nm) was selected for IDT metallization. Further, 100 nm $SiO_2$,[34] was deposited by magnetron sputtering on top of the wafer. This layer acts as a passivation layer and protects aluminum IDTs from corrosion. This $SiO_2$ thin film confers a hydrophilic character to the surface of prepared chips (wetting contact angle ca. 35°). The contact pads were opened for electrical contact by a dry etching procedure. **Figure S1** in the Supporting Information section (SI) shows the utilized mask design and a manufactured wafer aimed at fabricating IDTs with wavelengths varying from 120 µm to 510 µm and an aperture of 9.95 mm, including various designs with a different number of finger pairs. Chip sizes of 15 x 20 $mm^2$, 15 x 25 $mm^2$, and 15 x 30 $mm^2$ have been selected depending on wavelength. The devices with 120 µm and 510 µm wave length (henceforth called *120 chip and 510 chip*, respectively), used for the de-icing experiments of small ice aggregates carried out in this work, were prepared on 15 x 20 $mm^2$ and 15 x 30 $mm^2$ LN pieces, respectively. This ensured a minimum area of 15 x 10 $mm^2$ without IDTs for de-icing experiments. Chips operating at different wavelengths were diced into individual chips, as shown in **Figure S2** in the SI section. Keeping the same apertures and number of finger pairs, different IDTs layouts or sizes of diced LN plates were used for the chips utilized for proof of concept experiments, as explained in the corresponding section.

For the electroacoustic characterization of the chips as a function of finger pairs, on-wafer electrical measurements were carried out using a vector network analyzer (VNA, Keysight E5080B) and a wafer prober (**Figure S3** in the SI section). Before RF measurement, a viscous photoresist was applied between the IDTs for wave attenuation to neglect reflections on neighbor IDTs and thus avoid undesired effects by measuring the reflection coefficient ($S_{11}$) curves.

## 3.6. 2.2. De-icing and active anti-icing tests with small ice aggregates

De-icing and anti-icing experiments were carried out in a custom-made icing chamber, where temperatures down to a minimum value of -20 °C can be controlled (see **Figure S4**, showcasing a view of this chamber, the chip holder, and the water dosing system). An ad-hoc water dosing system consisting of heatable tubing inserted via the wall of the cooling chamber was employed to enable the dosing of droplets of bi-distilled water on the surface of the chips to produce ice particles. Inside the chamber, the tube was kept at a temperature of 2 °C to prevent droplet freezing inside the tube. The temperature of the substrate was controlled independently from that of the cooling chamber by a Peltier plate on which the holder and PCB were placed. Several thermocouples were mounted on the stage and the droplet hose to precisely control the temperature of the chip and water droplet dispenser. A USB camera with an attached C-mount objective was placed out of the cooling chamber to follow the evolution of ice aggregates and droplets.

De-icing and active anti-icing experiments were conducted on chips with IDTs of 120 µm and 510 µm in this custom-built cooling chamber. The icing tests were performed with the best-suited configuration of available chips (according to the number of finger pairs) as determined by their electro-acoustical characterization.

Experiments were carried out at three temperatures (-5 °C, -10 °C, and -15 °C, as controlled on the Peltier plate) with the stage in a horizontal position. At these temperatures, the porosity and nature of ice formed upon accretion processes from the impact of supercooled water droplets change from glaze (at -5°C) to mix (at -10 °C) to rime (at -15°C). However, under the experimental conditions of herein-reported experiments, the ice did not form by accretion but by cooling sessile water droplets. These conditions are prompt to generate little porosity. The stage temperature was continuously monitored during the experiments using a thermocouple mounted on an aluminum holder in direct contact with the Peltier plate. Once the temperature was stabilized, a 22 µl milliQ water droplet at +2 °C was dispensed on the surface of the chips at a distance of approximately 3 mm from the IDTs. After a while, the dispensed water droplet froze into a small aggregate of ice, and then the chip was electrically excited until the ice was completely melted. The minimum activation power (we denote this de-icing power as $DP_T$, a magnitude that depends on temperature T) has been taken as a relevant parameter for comparing the de-icing efficiency of the 120 µm and 510 µm chips. For the active anti-icing experiments, the same volume of water was dripped on an electro-acoustically activated chip, and the minimum power required to prevent freezing (anti-icing power, $AP_T$) was taken as the relevant parameter for comparing efficiencies. In these experiments, the surface was permanently activated to avoid ice nucleation and the subsequent freezing of the deposited water droplet.

The RF AW activation of the chips inside this cooling chamber was done by contacting the diced chips via a 50 Ω impedance printed circuit board (PCB) provided with gold-coated spring-pin connectors as shown in Figure S2 in the SI section together with the chip holder adapted to conduct the de-icing experiment. An RF supply, BSG F20 (BelektroniG GmbH, Germany), has been used for electroacoustic excitation, and the power supplied by the instrument is taken as a reference parameter of applied power.

**3.7.    2.3. Proof of concept: experiments for de-icing large ice layers with Lamb waves**

Two proof-of-concept experiments using Lamb waves have been carried out to determine the de-icing mechanisms of this type of wave applied to large-area ice layers or aggregates.

**Experiment 1**: In the 510 µm chip used for this experiment, the IDT and pad for connections were structured on the LN face that was not exposed to ice accretion. This means the 510 µm IDTs and PCB board were on the opposite side of the ice-formation surface. The connecting board was adequately protected to prevent shortage or malfunction with the IDT layout in the center of the LN plate (**Figure S5** in the SI section shows the chip, PCB, and protecting board). With this configuration, the icing area is similar to the 15 x 30 mm$^2$ chip size. Ice accretion was done in the IWT facility available at the TAU laboratory (detailed information about these Ice Laboratory facilities can be found in Koivuluoto et al.,[35]). A maximum wind speed of 25 m/s, with a liquid water content of 0.5 g/m$^3$ at a temperature of -10 ºC was used for the reported experiment. The formation of a mixed ice type is expected from these experimental conditions. An ice film with a thickness between 0.5 mm and 1 mm, depending on the location on the chip surface, was accreted and kept in the wind tunnel chamber at -10 ºC before being subjected to AW activation.

**Experiment 2**. These experiments were performed with a 510 µm chip of 15 x 30 mm$^2$ with the IDT structured on the ice-exposed side of the LN plate. In this case, a larger volume of water of approximately 180 µl was deposited in ambient air on the chip surface outside the zone of IDTs. The water covered an area of approximately 0.9 cm$^2$, separated by a distance of approximately 2 mm from the IDT. Then, the chip and connecting board were placed inside the cooling chamber, and the temperature decreased up to -15 ºC to induce rime ice formation. Afterward, the chip and board were removed from the cooling chamber, and the AW de-icing was induced in ambient conditions immediately after connecting with the electronics (total time shorter than 10 s). The process was followed by placing a video camera outside. During this time, the sample temperature never surpassed 0ºC without AW activation. These conditions generated some frost on top of the surface of the chip due to water vapor condensation, as visible during the experiments.

For the two experiments, the electronic control and activation systems were placed outside either the cooling chamber (experiment 2) or IWT (experiment 1, although in this case, the ambient temperature was -10 ºC). Electronics consisted of a VNA (vector network analyzer) to determine the working frequency during and after ice accretion and a switch allowing to automatically connect the chip to a signal generator system consisting of a signal generator (Keysight33210A) and an amplifier (Mini-Circuits LZY-22+). Further details of this electronic control and activation system can be found in Del Moral et al.[24] The reported de-icing experiments were carried out applying an RF signal at around 7.3 MHz with slight variations in the order tenths of kHz depending on the chip and peak-to-peak voltages of 53 V (experiment 1) and 70 V (experiment 2), equivalent to approximate power values of 1.7 W and 4.8 W, respectively.

### 3.8.  2.4. Finite Element Model (FEM) calculations

Finite element (FE) simulation of the 120 µm and 510 µm chips was conducted using COMSOL Multiphysics version 6.0. This simulation describes the interaction between AWs and ice, verifying the easiness of mechanical energy transmission through the piezoelectric LN and the ice interface. For this purpose, we have calculated the displacement field ($u$) in the substrate and the ice aggregate under the assumption that larger displacements in the ice aggregate mean a better transmission of the mechanical energy of the AW from the activated LN. The ice aggregate was located at the same distance from the edge of the last finger of the IDT as for the experiments, trying to reproduce the actual conditions of the conducted experiments. To keep the computational costs within reasonable limits, the volume of the simulated droplet has been settled at 4 µl. Ice aggregates were assumed to present a hemispherical shape, as it typically happens after freezing water droplets deposited on surfaces.

The AW vibrations of the LN plate with IDTs have been modeled using a series of electrodynamic and mechanic equations, as previously described by Fakhfouri et al.[36] Since R-SAW and Lamb waves on 128° X–Y LiNbO$_3$ are mainly polarized in the median plane,[37] a two-dimensional (2D) model was simulated using a generalized plane strain assumption. This assumption has been adopted to simulate both the 120 µm and 510 µm chips. The 0.5 mm thick substrate has been described with the elasticity, piezoelectric, and electric permittivity tensors of LN,[38] which were adequately rotated to account for the actual 128° X–Y crystal cut of the substrates used in the experiments. The ice was modeled as an isotropic linear-elastic material in the form of a semicircle geometry, using the ice properties provided by Victor and Whitworth[39] that describe a compact type of ice similar to glaze ice. All this information, including the description of the characteristics of the IDTs, dimensions, applied voltage, time step used for the simulations and dimension, and other computational parameters, can be found in the SI section, **SI6.** Notably, the ice-substrate interface between the ice and the LN substrate has been simulated as flat, an assumption that fits well with the conditions of the real system formed by a single crystalline substrate with very little surface roughness. An ideal system consisting of an R-SAW with a long wavelength (510 µm) interacting with the ice aggregate was also simulated for comparative and discussion purposes. The parameters used for computing the 510 µm Lamb wave system were kept for this ideal system, except for the assumption of a finite thickness for the LN substrate. This semi-infinite thickness boundary condition precludes any reflection of the AW at the bottom face of the plate, i.e., imposing a situation typical of the generation of R-SAWs.

**Figure 1** presents a scheme with the definition of system coordinates (x, y, z) used for calculations. Since the COMSOL Multiphysics calculations have been carried out in two dimensions (2D), the relevant coordinates are x and z (c.f. Figure 1b). The former corresponds to the propagation direction of the wave, while the latter corresponds to the thickness of the LN plate.

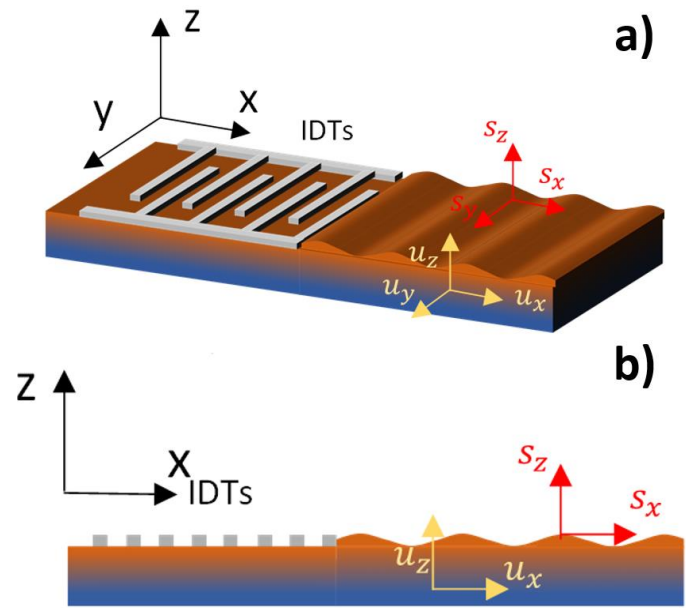

*Figure 1.- Schematic description of the geometry of chips with displacement field and axis definitions; a) 3D diagram of the chip; b) 2D diagram of the chips as used for the calculations*

For calculations, figures, and data presentation, we have adopted the following definitions:

*Volumetric displacements*: the local, wave-induced displacement field undergone by the lattice of the LN plates. They are characterized by a lattice displacement vector **u**. Relevant magnitudes for data calculation and presentation are its modulus in 2D |**u**| and the two components $u_x$ and $u_z$.

*Surface displacements*: the displacement field undergone by the outer surface of the plate, which in the icing experiment is in contact with the ice and is directly involved in the transmission of the mechanical energy to the ice. This definition simplifies the notation and remarks on the importance of power confinement on the surface of the chip. Surface displacements are characterized by a surface displacement vector, **s**. Relevant magnitudes to present data and for calculations are its modulus in 2D |**s**| and the two components $s_x$ and $s_z$.

The geometry of the whole system has been discretized with an unstructured triangular mesh, and the acoustic modes have been described with at least ten elements per wavelength considering the following sound velocities: in LN substrate 3981 m/s,[38] in ice 1850m/s.[39] The resulting grid had approximately three million elements (equivalent to approximately 18.3 million degrees of freedom). Electrical excitation was modeled using an electric potential that follows the periodicity of the IDT fingers, which were located on the left half of the model geometry, neglecting the mechanical damping of the metal electrodes. To avoid undesired reflections of the AWs at the edges of the chips, the AWs were absorbed by incorporating "perfectly matched layers" (PML) as defined in the COMSOL software documentation.[40]

The model was solved in both the frequency domain and the time domain using a frequency solver or a time-dependent solver, respectively. In the frequency domain, the IDTs were defined as an RF port where power is explicitly defined as an input parameter (1 W was arbitrarily selected for the calculations since simulations are linear and provide $S_{11}$ spectra and harmonic solutions of the displacement and electric fields, no dependence of results is expected on the used input power). Simulations were also done in the time domain to obtain real-time estimation of the vibroacoustic response of the LN plate and the LN/ice systems upon electroacoustic excitation. In this case, the electrical excitation is applied to the IDTs as a sinusoidal voltage signal with defined amplitude and working frequency as input parameters. Working frequencies were selected based on the frequency domain results following maximum power transmission criteria. To generate equivalent mechanical powers for the AWs in the simulation of the 120 µm and 510 µm chips, the selected voltage amplitudes in each case were different to compensate for the differences in electroacoustic coupling. Therefore, the resulting electric power for each wavelength was slightly different to ensure that the generated mechanical power was equivalent in all simulated models.

## 3.9. 3. Results and Discussion

### 3.10. 3.1. Electrical characterization of Acoustic Wave chips

The radio-frequency electrical characterization of the chips was performed by measuring the $S_{11}$ (return loss) and $|S_{11}|^2$ (reflection coefficient of power) parameters as a function of frequency. The measured resonance frequencies (frequency of the minimum of the $S_{11}$ peaks) closely match the calculated resonance frequencies of the chips, according to the well-known expression:

$$f = v_s/\lambda \tag{1}$$

where $f$, $v_s$ and $\lambda$ are, respectively, the working frequency, the wave velocity on 128º YX LN substrate (3981 ms$^{-1}$ [38]), and the wavelength, this latter defined by the IDT layout (i.e., the distance between fingers). The series of return loss spectra plotted as a function of frequency recorded for the different chips prepared in this work are reported as **SI7**. These spectra show that the $S_{11}$ minima deepen for a higher number of finger pairs. To minimize the reflected power, de-icing tests were carried out with 120 and 510 µm chips with the largest number of finger pairs tested during the electroacoustic evaluation, i.e., 15 and 26, respectively.

To estimate the power efficiency, the reflection coefficient of power $|S_{11}|^2$ is the relevant parameter, which can be calculated from equations (2) and (3):

$$RL(dB) = 10 \, \log_{10}\left(\frac{P_i}{P_r}\right) \tag{2}$$

$$|S_{11}|^2 = \frac{P_r}{P_i} \tag{3}$$

where *RL(dB)* is the return loss of the chips and $P_r$ and $P_i$ are reflected and incident power, respectively.

**Figure 2** shows a series of plots of the reflection coefficient of power $|S_{11}|^2$ versus frequency as recorded for the 120 µm and 510 µm chips with different numbers of finger pairs. A photograph of these two chips is also included in the figure. The full set of $|S_{11}|^2$ curves recorded for all manufactured chips is shown in **Figure S8**.

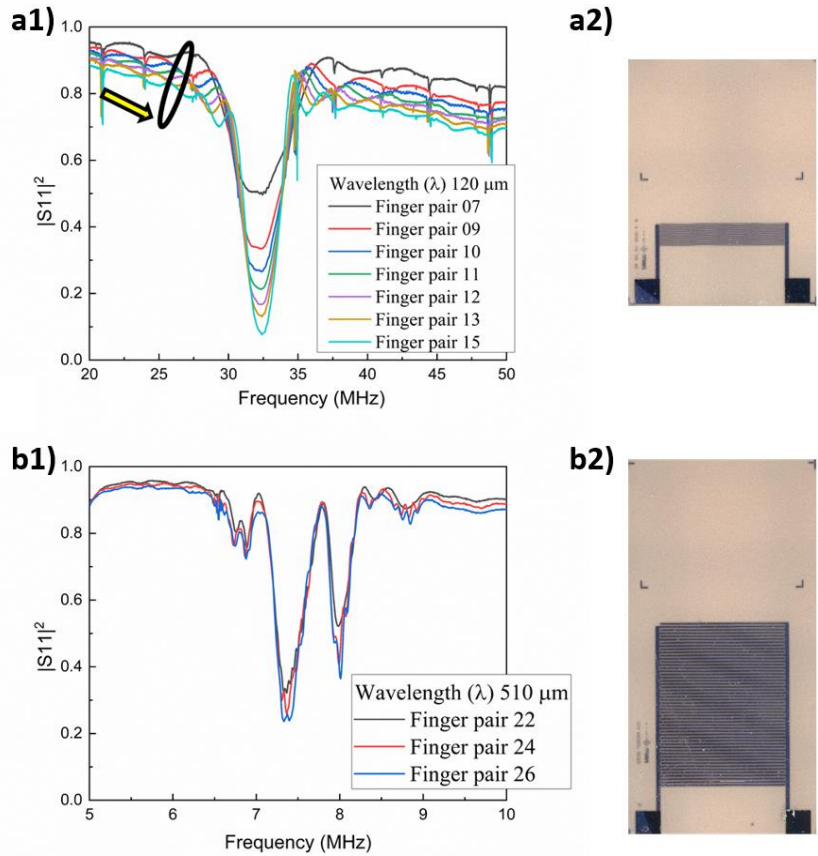

**Figure 2. Electrical characterization of AW chips:** a1) and b1) $|S_{11}|^2$ versus frequency plots for fabricated AW devices with wavelengths (λ) of 120 µm (a1) and 510 µm (b1). Note the different spam of the x scales used in each case. The arrow and ellipse highlight the progressive increase of power losses in the series of 120 µm devices as the number of finger pairs increases. Photographs of the 120 µm (a2) and 510 µm (b2) chip devices with the maximum number of finger pairs.

The plots in Figure 2 show that the reflected power decreases when the number of finger pairs increases. For the longest wavelength chip, i.e., the 510 µm chip, power losses reached saturation at $|S_{11}|^2$ values around 0.25 for 26 finger pairs. In general, although the reflection coefficient of power could still slightly decrease by increasing the number of finger pairs, the benefits would be very small and likely compensated by the electrical power losses that also increase with this parameter. Therefore, we have selected 15 and 26 finger pairs for the 120 µm and 510 µm chips for de-icing and anti-icing experiments. This choice is a good compromise between low reflection and an acceptable free chip area for de-icing experiments (see below). These plots reveal a lowering in baseline (ideally, $|S_{11}|^2$ should equal one outside of the resonance peak), which is particularly noticeable for the 120 µm chip. This indicates the existence of some electrical power losses, which are not due to the resonance excitation of the crystal and are likely caused by factors such as parasitic capacitances in the IDT, ohmic resistance of the thin film metallization, or contact resistances. These parasitic losses are in the order of 10-20% for the characterized chips, a percentage of the incident power that will not contribute to the generation of AW vibrations through electromechanical coupling. The plots in Figure 2 also show that, as expected, these electrical power losses increase with the number of finger pairs.

According to Figure 2 and Figures S7/S8 in the SI section, the electroacoustic response of the 510 µm chip is different. The behavior of this chip is characterized by two resonance minima instead of one in the frequency region of interest. We attribute these two resonances to Lamb plate waves instead of Rayleigh waves. According to the well-established theory of SAW generation in piezoelectric plates,[41] an R-SAW can be regarded as the high-frequency limit of the superposition of symmetric and anti-symmetric Lamb waves. These two counterparts start to separate for plate thicknesses of around 1.5 $\lambda$, when the Lamb wave velocities are close to SAW velocities. From the $|S_{11}|^2$ plots in Figure 2, it appears that when the wavelength of the AW approaches the thickness of the plate, the conditions for pure Rayleigh-SAW are no longer fulfilled. According to these considerations and Equation 1, the resonance at 7.36 MHz corresponds to the A0 Lamb wave mode, whereas the ~ 8 MHz resonance must be attributed to the S0 Lamb wave mode. In the next section, we further support this attribution by the simulations obtained using COMSOL Multiphysics analysis of the AWs in the 510 µm chip. It is noteworthy that chips with a similar configuration but a wavelength larger than 510 µm behave similarly and generate two AWs at different frequencies (see **Figure S9** in the SI section for the $|S_{11}|^2$ curves experimentally determined for a 600 µm chip, which is reported for comparison).

**Table 1** includes experimental and calculated (i.e., Equation 1) resonance frequencies for the manufactured chips with the maximum number of finger pairs in each case, as well as the type of generated AW.

**Table 1**. Calculated and experimental frequencies for the best IDT configuration of the chips in terms of the number of finger pairs.

| Wavelength (µm) | Number of finger pair | Measured Frequency (MHz) | Calculated frequency (MHz) | Mode |
|---|---|---|---|---|
| 120 | 15 | 32.32 | 33.25 | R-SAW |
| 150 | 16 | 25.79 | 26.60 | R-SAW |
| 240 | 20 | 16.10 | 16.62 | R-SAW |
| 330 | 22 | 11.68 | 12.09 | R-SAW |
| 420 | 28 | ~ 9.20 | 9.50 | R-SAW |
| 510 | 26 | ~ 7.36 & 8.14 | - | Lamb wave |

### 3.11. 3.2. FE simulation of acoustic waves

Simulation results of the AW generation and propagation in the 120 µm and 510 µm chips were carried out as described in the Experimental and Methods section. **Figures 3 a)** and **b)** present selected results of this simulation analysis for each type of wave. The videos provided as SI section **S10-S12** showcase the time-dependent variation of volumetric displacements represented in the form of color maps with the same color code as for the snapshots in Figures 3 a2) and b2).

The plots in Figure 3 a1) and b1) compare in the frequency domain the experimental and calculated $|S_{11}|^2$ spectra of the 120 µm and 510 µm chips. The concordance between curves obtained with experiments and simulations confirms that simulations reproduce the experimentally recorded signals, except for the electrical IDT power losses evidenced by the non-zero backgrounds of the experimental curves, a feature particularly noticeable for the 120 µm R-SAW. This proves that, except for this deviation attributable to ohmic losses and/or capacitive effects in the real chips not included in the analysis, the simulations for the two chips agree with their actual behavior.

The analysis in Figure 3 a2) of the volumetric displacements determined in the time domain for the 120 µm chip confirms that vibrations only affect the outer surface layers of the LN plate, as expected for a typical R-SAW type (see video S11 in the SI section). This is evidenced by the color maps in this figure showcasing snapshots of the modulus ($|\mathbf{u}|$) and $u_x$ and $u_z$ components of volumetric displacements of this wave that confirm the confinement in the outer surface regions of the plate. The plots in the same figure representing the value of $|\mathbf{u}|$ as a function of the thickness coordinate z show that the displacement field goes to zero in the bulk.

Figure 3 a3) presents the plots of surface displacement components $s_x$, $s_z$, and the modulus of the displacement vector at the surface $|s|$ along the direction of wave propagation (x). Notably, these parameters correspond to the AW-induced displacements in the plate outer plane of the chip plate, i.e., the surface that will be in contact with the ice. Figure 3 a3) reveals that maximum displacements along x and z are almost equivalent and are separated by a phase shift of 93º. This result means that the surface displacement field approaches a circular polarization, as shown by the plot in Figure 3c) representing the component $s_x$ vs. $s_z$ (note that the different circles in the plot refer to calculations at different times).

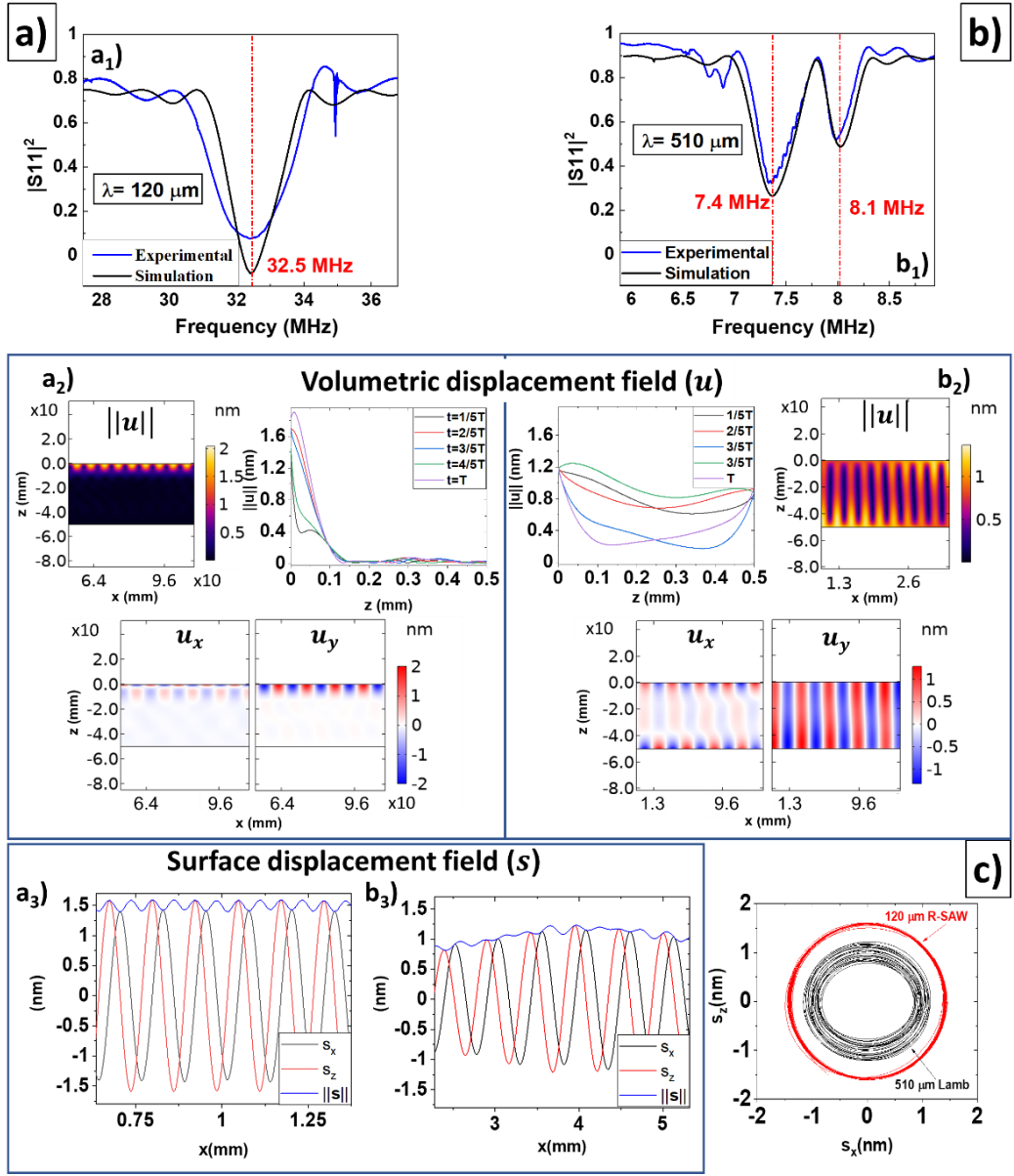

**Figure 3. Simulation of the AWs:** a) and b) simulations for the for the 120 µm (a) and 510 µm (7.4 MHz peak) (b) chips. a1 and b1 simulation in the frequency domains of $|S_{11}|^2$. a2 and b2 simulation in the time domain of bulk displacements. Simulations include: i) color maps snapshots of bulk displacements (modulus |**u**| and $u_x$ and $u_z$ components); ii) plots at various times of the modulus of bulk displacement vector |**u**| along the coordinate z (thickness of the plate). a3 and b3 plots the modulus |s| and sx and sz components along coordinate x at a given time. Note that the x scale is not the same for the two waves to properly compare the displacement fields. c) Plot of $s_x$ vs $s_z$ components to assess the polarization of the displacement field at the surface in the two chips.

On the other hand, the time domain simulation of the electroacoustic activation of the 510 µm chip in Figure 3 b1 gives rise to two well-defined and distinguishable $|S_{11}|^2$ peaks matching the shape of the experimental curves, as well as the frequency values of the minima (Table 1).

Meanwhile, the analysis in Figure 3 b1) demonstrates that volumetric displacements affect the whole plate thickness. The color plots in this figure show that the first peak at 7.4 MHz (note that there is a small difference with respect to the experimental minimum at 7.36 MHz) corresponds to vibrations with anti-symmetrical character and should be attributed to a A0 vibrational mode. Meanwhile, volumetric displacements corresponding to the second peak at 8.1 MHz have a symmetrical character (i.e., typical of an S0 mode, data not presented). Meanwhile, plots in this figure of |u| as a function of the coordinate z within the plate show that the displacement field does not vanish in its interior and extends from one face to the opposite of the LN plate. Figure 3 b3) shows that the magnitude of $s_x$ and $s_z$ components is similar, and there is a phase shift of 91° between them, i.e., the surface displacements are circularly polarized. This is confirmed by the representation of $s_x$ vs $s_z$ in Figure 3c).

When comparing the waves generated in the 120 µm and 510 µm chips, it is noteworthy that they present surface and plate character, respectively. Another significant difference is that the surface displacements in the 510 µm chip for the A0 mode are slightly smaller than for the 120 µm chip, and they present a certain modulation due to the superposition of a second contribution of much larger wavelength in the order of 4-5 mm. This second contribution is likely due to a weak standing wave mode with a larger wavelength that is superimposed on the main one (see **Figure S13** in the SI Section).

### 3.12. De-icing experiments of small ice aggregates

To verify the effect of the AW on the de-icing process, we have systematically compared a series of de-icing experiments carried out with the 120 µm and 510 µm chips. Experiments were done at temperatures of -5, -10 and -15 ºC. A summary of the experimental conditions is shown in Table 2, including information about temperature, actual frequency, and incident power values (i.e., DPT values), and an estimation of the effective power actually available for AW excitation. The latter have been calculated after the correction of the experimental $DP_T$ values. A first correction considers that the electrical power that is not inserted into the LN plate in the form of AW electromechanical activation would be equivalent to 1- $|S_{11}|^2$ (for the 120 µm chip with 15 finger pairs, it appears that approximately a 8% of the incident electrical power is not efficiently coupled to the device). The already mentioned parasitic electrical power losses associated with the IDTs should be summed up in this correction. Parasitic losses were important for the 120 µm chip with 15 finger pairs, which amounted to ca. 20% of incident power (c.f., Figure 2). In total, a rough estimate of the electrical power actually converted into mechanical excitation with this chip might be in the order of 72% of the incident power (this figure stems upon substrating the 20% of parasitic lossess plus ca. 8% due to the coupling issues). For the 510 µm chip, a similar analysis gives an estimated correcting factor of 69% (note that in this case ohmic or capacitive losses associated to the IDTs were smaller, c.f. Figure 2). Since the estimated power losses resulting from these correcting factors were rather similar for the 120 µm and 510 µm chips, comparative assessments of powers and efficiencies would be equivalent using either incident or effective power values. Note, however, that heating effects due to ohmic losses at the IDTs will likely be higher for the 120 µm chip.

**Table 2.** De-icing data for the 120 µm and 510 µm chips: temperature, operating frequencies, incident ($DP_T$), and effective AW powers after correction by the estimated losses. The uncertainty limits for the $DP_T$ values define the variation in the measurements in each case.

| Chip (µm) | Temperature (ºC) | Frequency (MHz) | Incident power ($DP_T$) (W) ± error | Effective power (W) |
|---|---|---|---|---|
| 120 | -15 | 32.33 | 3.76 ± 0.26 | 2.70 |
|  | -10 | 32.40 | 2.76 ± 0.68 | 1.98 |
|  | -5 | 32.44 | 1.46 ± 0.20 | 1.05 |
| 510 | -15 | 7.36 | 2.02 ± 0.39 | 1.39 |
|  | -10 | 7.40 | 1.46 ± 0.29 | 1.00 |
|  | -5 | 7.36 | 1.45 ± 0.16 | 1.00 |

De-icing experiments proceeded in the following way: once the chip temperature was stabilized in the cooling chamber, the system was kept under these conditions for at least 30 minutes for temperature homogenization. Then, a liquid droplet (+2 ºC, 22 µl) was delivered and frozen on the surface of the chips, and an RF signal with the incident ($DP_T$) power reported in Table 2 was applied to generate AWs at the resonance frequencies of each chip. Reported values correspond to the minimum incident power required to induce melting. Both $DP_T$ and effective AW power values in Table 2 confirm that melting on the 510 µm chip is more favorable than on the 120 µm chip. We tentatively attribute this difference to the distinct characteristics of the waves generated in each case, a feature that will be further discussed in the following sections.

The melting sequence induced upon application of the reported powers in **Table 2** was video-recorded for both chips. **Figure 4** shows a series of snapshots representing dripping (Figure 4 a1 and b1), freezing (Figure 4 a2 and b2), partial melting (Figure 4 a3 and b3), and final melting (Figures 4 a4 and b4) stages of the experiments performed at -15 ºC. Results of similar experiments at -5 ºC and -10 ºC are reported as SI section in **Figure S14** in the SI section and the videos in **S15** and **S16**.

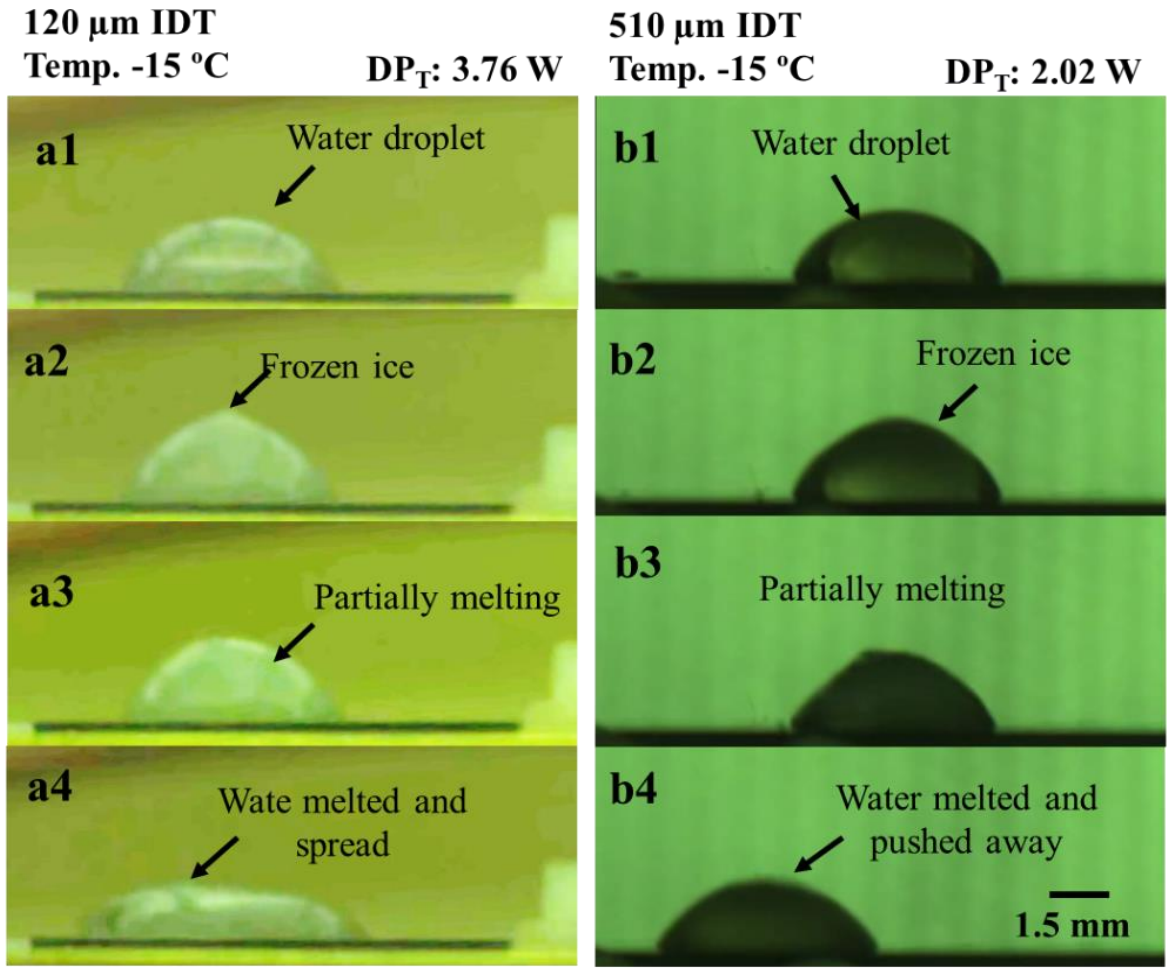

**Figure 4. Comparison of de-icing process on the 120 µm (a) and 510 µm (b) chips.** A 22 µL droplet is delivered on the chips at –15 °C (a1, b1); droplet is frozen; (a2, b2): droplet starts melting, after chip activation with the $DP_T$ powers reported in table 2 (a3, b3); droplet remains liquid (active anti-icing effect) (a4, b4). The behaviour of the melted water droplet was different for the two chips: it spread for the 120 µm chip but was pushed toward the chip edge in the case of the 510 µm chip. The water droplet was dispensed at almost similar distance from the last finger of the IDTs (i.e., at about 3 mm).

In general, longer times were required for de-icing with the 120 µm than the 510 µm chips. For example, at -5 °C only a slightly higher power of 1.46 W was required for de-icing with the 120 µm chip against 1.45 W for the 510 µm chip (c.f. Table 2). However, the droplet began to melt after ~70 seconds of activation and completed the process after 170 seconds on the 510 µm chip, requiring 230 seconds to achieve complete melting with the 120 µm chip. A similar behavior was also observed at the other temperatures. Thus, at -15 °C, the power required to melt the ice aggregate was 3.76 W with the 120 µm chip and applied for a time longer than 215 seconds for completion. Remarkably, at this temperature, less power was needed with the 510 µm chip, for which an incident power of 2.02 W and an excitation time of 65 seconds was sufficient to achieve complete de-icing.

Additional differences were observed during the melting process and in the final melted states. In the intermediate states (i.e., stages a3 and b3 in Figure 4), the droplet shape became occasionally asymmetric on the 120 µm chip, while it remained always symmetric on the 510 µm chip. This occasional asymmetry might be a hint that melting has begun on the right side of the droplet, where liquid water would form due to the interaction with the traveling R-SAW moving from the right to the left in the images. Such a lateral melting mechanism has been previously discussed by us in Jacob et al.,[25] dealing with R-SAWs. Meanwhile, the preservation of the symmetric shape of the droplets on the 510 µm chip suggests that melting is progressing through the whole ice-substrate interface with no lateral inhomogeneity in the water-ice distribution. Significantly, once the droplets are entirely melted (stages a4 and b4 in Figure 4), strong vibrations were visible inside the ice drop with the 510 µm chip, the droplets displacing further toward the edge of the chip. Water displacements have been amply reported in the literature for the AW excitation of water droplets.[42-44] The use of AW to induce de-fogging, removal of water droplets, or cleaning of surfaces has also been studied.[45, 46] In these experiments (c.f. Figure 4 and Figure S14 in the SI section), water droplet displacements were not observed on the 120 µm chip. Thus, water droplets spread laterally in the direction of propagation of the wave (i.e., their contact angles decreased) while keeping their original position with respect to the IDTs. From these experiments, we conclude that ice melting and water droplet removal from the surface was more favorable for the 510 µm chip, i.e., the chip operating with long wavelength Lamb waves.

### 3.13. Active anti-icing of water droplets

In real-world scenarios, avoiding ice accretion on the surfaces of substrates exposed to icing conditions is an essential function. We call this function active anti-icing to explicitly distinguish it from using anti-icing coatings extensively utilized for passive systems.[1-5] To test the active anti-icing capability, the chips were continuously activated with the minimum power required to prevent the transformation of water into ice under environmental conditions that otherwise induce the freezing of water droplets. It should be mentioned that the $AP_T$ values determined in this way may be affected by inaccuracies inherent to the determination of this minimum power and that for discussion, data should be considered only in a semi-quantitative way.

In the experiments, a water droplet (22 µL, at 2 °C) was delivered onto the surface of the RF-activated chips kept at the temperatures gathered in **Table 3**. Activation frequencies, incident (denoted $AP_T$), and effective powers are also included in this table. To estimate the active anti-icing power, various dripping trials were done with the chip excited at increasing incident powers until the water droplet remained liquid on the surface. Experiments were repeated several times to determine the reproducibility of the tests. It is important to stress the dynamic character of the experiment and that slight differences in droplet impact on the surface or surface location may affect the value of the minimum incident power required to keep the water liquid. Hence, the values of incident power in Table 3 were obtained by averaging the results for the minimum values obtained in at least three experiments where freezing was not observed. A preliminary assessment of the reported incident (i.e., $AP_T$) and effective power values in this table in comparison with data in Table 2 reveals that, for a given temperature, $AP_T$ was always smaller than $DP_T$. The results in Table 3 also highlight that $AP_T$ decreases when temperature increases. Significantly, it also indicates that the threshold incident power in active anti-icing experiments was always smaller for the 510 µm chip.

**Table 3.** Active anti-icing results for the 120 µm and 510 µm chips: temperature, incident and effective powers, and operating frequencies. All experiments were carried out for 22 µL droplets with the nozzle temperature at 2 ºC. The uncertainty limits for the $AP_T$ values define the variation in the measurements in each case.

| Chip (µm) | Temperature (ºC) | Frequency (MHz) | Incident power $AP_T$ (W) | Effective power (W) |
|---|---|---|---|---|
| 120 | -15 | 32.24 | 3.99 ± 0.01 | 2.87 |
|  | -10 | 32.34 | 2.08 ± 1.0 | 1.49 |
|  | -5 | 32.16 | 1.46 ± 0.32 | 1.05 |
| 510 | -15 | 7.40 | 1.42 ± 0.13 | 0.98 |
|  | -10 | 7.39 | 1.29 ± 0.25 | 0.89 |
|  | -5 | 7.36 | 1.06 ± 0.01 | 0.73 |

### 3.14. Simulation of AW interaction with small ice aggregates

To shed some light on how the AW-ice interaction affects the de-icing mechanisms on the two devices, we have simulated how the mechanical energy of the waves is transmitted to compact droplets of ice placed on the flat surface of the LN substrate. During ice accretion processes (e.g., on the surface of airplane wings), the formation of compact ice is expected when glaze ice is formed at around -5 ºC, but not when ice forms by accretion or frosting at temperatures of the order of -15 ºC or lower, a situation by which the formed rime ice uses to be relatively porous. Selected simulations at different times reported in **Figure 5** (see videos of the complete series of calculated events in **S17** and **S18**) showcase a series of time-dependent snapshots of the modulus of the volumetric displacements generated both in the plate and the ice aggregate surface by the propagation of the AWs for the 120 µm R-SAW (Figure 5 a) and for the 510 µm Lamb wave (Figure 5b). Furthermore, Figures 5 c) and d) represent the evolution of amplitude of the surface vibrations in the outer plane of the LN plate along the wave propagation direction before and once the AWs have entered the ice-substrate interface. Similar simulation results are presented for the 510 µm R-SAW in **Figure S19** and video **S20** in the SI section**.** At this point, it is noteworthy that the presented snapshots, plots, and videos stemming from the simulations do not aim to describe the melting process but to assess how and in which zone of the ice-LN interface the AW mechanical vibrations transmit their energy to ice. Therefore, the reported simulations would not have any real physical significance whenever, in real conditions, a small amount of water is formed at the interface or the edge of the ice aggregate due to the accumulation of energy in this zone of the ice droplet.

The snapshots in Figure 5 a) and the plot of the amplitude of surface oscillations in Figure 5 c) illustrate that the 120 µm R-SAW experiences an effective and complete damping within an interface zone of approximately 10-11 wavelengths (i.e., approximately 1 mm). These simulations agree with previous results reported in Jacob et al.[25] Meanwhile, snapshots and plots of the surface amplitude of the 510 µm Lamb wave in Figures 5 b) and d) reveal that the amplitude of the wave is less affected in this case (i.e., damping is less effective) by the interaction with ice, and therefore, the wave is not fully attenuated behind the ice droplet. This means that Lamb waves can propagate their energy along larger zones of the ice-substrate interface than R-SAW.

Of particular relevance is the comparison of the 510 µm Lamb wave with the simulations done for the 510 µm R-SAW (see Figure S19 in the SI section). It is apparent from this comparison that although the latter can effectively traverse through the entire ice-substrate interface without experiencing complete damping, the attenuation percentage after crossing the ice droplet is in the order of 75% for the 510 µm R-SAW and just 35% for the 510 µm Lamb wave. Although this simulation analysis must be taken as preliminary and should be complemented with experimental evidences, it suggests that both wavelength and type of wave are essential characteristics affecting the AW-ice interaction.

The assessment of experimental data for the melting of water droplets reported in the previous section and the commented features from the simulations permit advancing some specific considerations about the de-icing mechanism:

1.- The 120 µm R-SAW is effectively damped when penetrating the ice-substrate interface, and its intensity almost decreases to zero at about half the substrate length covered by the small ice aggregate (Figures 5a and 5c). In other words, most energy carried by the AWs is effectively delivered to the ice within a short distance inside the ice edge zone in front of the IDTs. Under these conditions, it is most likely that the left side of the droplet facing the IDT will be converted into water. Once water starts to form, the physical model in Figure 5 a) loses the experimental meaning because it does not represent the mixture water/ice existing in the experiments and the fact that once water is formed, the transmission of mechanical energy from the R-SAW to the liquid phase will be more favorable than to solid ice (i.e., damping would be even more effective). This situation was already highlighted in our previous work describing the ice melting mechanism induced by R-SAW, where we proposed that once a first portion of liquid water forms on the side of the ice aggregate facing the R-SAW propagation front, a water heat streaming mechanism is the main responsible for de-icing.[25]

2.- The Lamb wave generated in the 510 µm chip is less damped along the ice-substrate interface and penetrates long distances along the surface covered with ice (Figures 5c and 5d). It is interesting to highlight that the simulations showcase that some but not complete damping occurs within the ice-substrate interface for the small ice aggregate used for the simulations. It can be realized that this limited damping makes that the whole ice-substrate interface of the ice aggregate is affected by the AW activation. Since the transmitted mechanical energy is not very high, the wave is expected to initially provoke ice cracking and surface softening along the entire ice-substrate interface before that melting takes place in a second instance. The next section provides additional experimental evidence of this mechanism. This analysis and the previous comparison in Figure S19 of the SI section between the simulations for the 510 µm R-SAW and 510 µm Lamb waves also support that both wavelength and type of wave (R-SAW or Lamb waves) are factors modulating the interaction mechanism between ice aggregates and AWs.

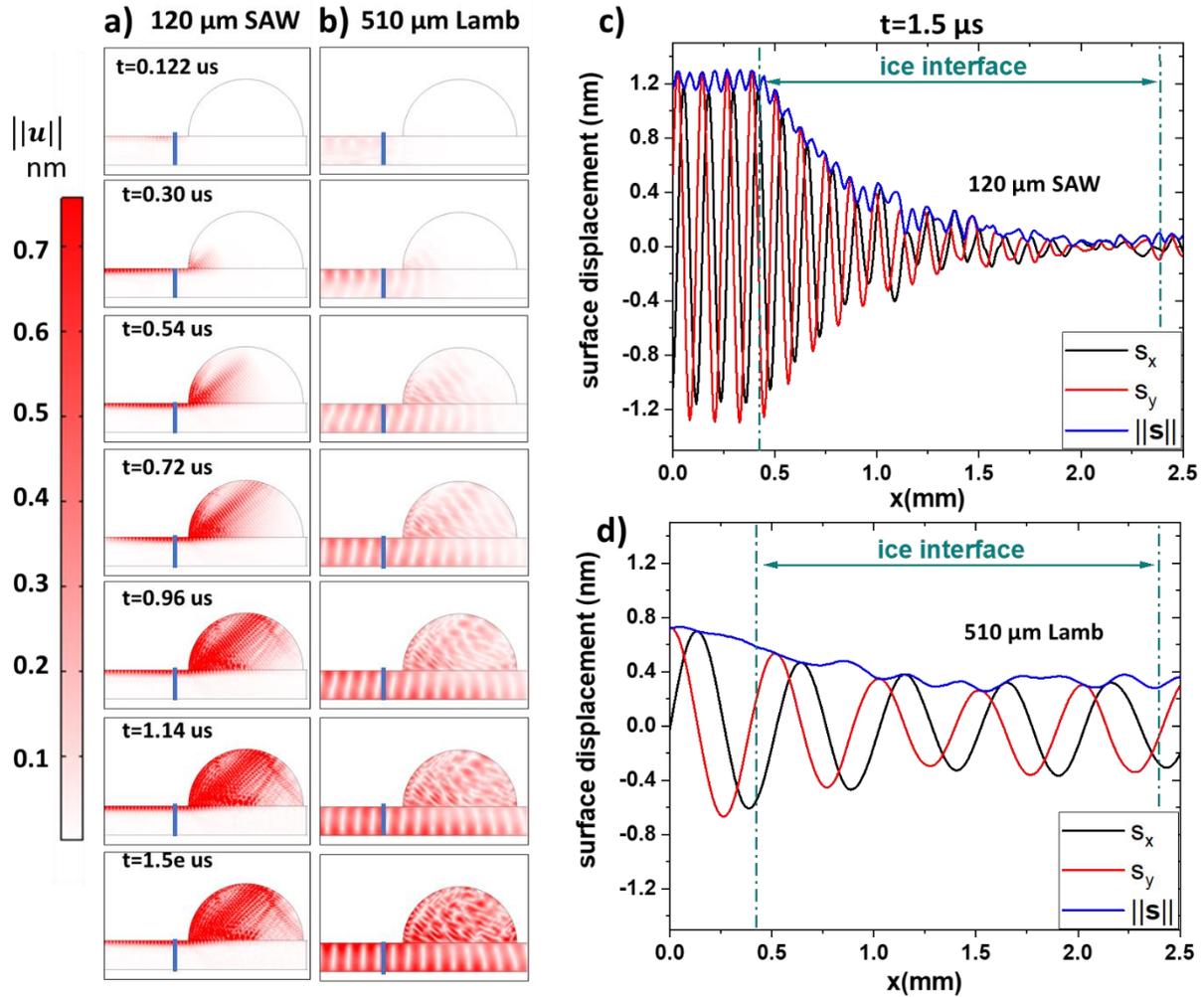

**Figure 5. AW-ice interaction and surface displacements.** Time-dependent simulated snapshots in the form of intensity color maps of the AW-induced displacements in the LN chips with an ice aggregate on their surface. The vertical line along the plate signals the location of the last finger of the simulated IDT. a) and b) Calculations for the 120 µm SAW and 510 µm Lamb wave, correspondently. c-d) Surface displacements at the LN substrate as a function of wave propagation along the x coordinate direction for the 120 µm SAW (c) and 510 µm Lamb wave (d) interacting with the ice aggregate on the surface.

### 3.15. Proof of concept de-icing of large ice aggregates and layers. De-icing mechanism

From the previous analysis and simulations, some key features can be deduced regarding the de-icing mechanism of small droplets: i) longer wavelength seems more efficient for ice melting (for the given droplet size); ii) melting with a traveling R-SAW is rather directional from one droplet side to the other and proceeds first in the edge facing the IDTs and iii) melting with a long wavelength Lamb wave is less directional due to a lower attenuation by the droplet and the reflection of the wave on the free side of the plate.

In a leap forward, in this section, we present two proof of concept de-icing experiments of large ice aggregates. Experiments were conducted with 510 µm Lamb waves and two different chip configurations. The purpose is to approach the situation encountered in real-world icing processes, where large amounts of ice will cover most, if not the complete, activated substrate. In Experiment 1 (see a detailed description in the Methods section), an ice layer with a thickness estimated between 0.5- and 1 mm became accreted in an open IWT at -10 ºC, conditions favoring the formation of mixed ice and some incipient porosity. AW de-icing was then activated in a cool room at -10 ºC. Purposely, the ice was accreted on the LN surface opposite to the IDTs. The IDTs we prepared in the center of the LN chip with the connected pads located at the two sides as indicated in Figure S5. This experimental configuration is intended to prove two features: first, that Lamb wave oscillations extend through the whole piezoelectric plate and, second, that these waves can induce de-icing on the surface opposite to the surface with IDTs (i.e., to prove that Lamb waves are effective on the two sides of the plate). A scheme of the experimental setup is reported as an insert in Figure 6 a). The evolution of the de-icing process in this experiment can be discussed following the series of snapshot photographs reported in this figure (see the original video in the SI section **S21)**. The first evidence is that chip activation enabled the effective removal of the ice accreted on the side opposite the IDTs and that the surface became free from ice. This proved that lattice oscillations extending through the whole thickness bear enough power to activate the de-icing process. Deicing took ~98 seconds from the application of the RF signal (photograph a1) up to the end of the process (photograph a7). Interestingly, de-icing involved a first step (photographs a2/a3) where ice cracking extended along the whole surface in about 10-12 s, from t = 36 s to t = 48 s. Melting occurred as a second step and took much longer than the initial cracking (about 50 s) from t = 48 s to t = 98 s. At the end of the melting process, the on-chip temperature was 10 °C, while in the close vicinity of the chip, the temperature remained at - 8 °C.

For clarity, the aforementioned de-icing steps are illustrated with a cartoon in the schematic included in Figure 5 a). We hypothesize that the initial cracking step in Schematic a3 is induced by the interaction of the ice layer with the activated substrate through the exchange of mechanical energy. The subsequent melting process in schematic a7 will mainly involve heating by acoustic-thermal effects and possibly some heat from the Joule effect due to ohmic losses at the IDTs beneath.

In Experiment 2, a larger aggregate of ice at -15 ºC is placed on the same surface as the IDTs**.** Since the ice did not form by accretion but by cooling a big droplet of previously deposited water, no significant porosity is expected, even if grown at this low temperature. The de-icing process took place according to a series of steps characterized by the snapshot photographs in Figure 6 b) (the video of the entire process is reported in the SI section **S22**) and illustrated by the schematics inserted in this figure**.** It is apparent that after 15 s of activation, the frost accumulated around the big aggregate of ice during the manipulation of the probe device disappeared. At this stage, the shape of the big aggregate of ice has not been modified, although changes in its aspect (brightness and whiteness of the aggregate) suggest that the interface with the substrate might be affected. Changes were more profound after 17 s and 22 s (snapshots b4/b5) of excitation when a mixture of ice/water coexists in the whole volume. A significant feature of this experiment is that no defined water/ice front formed during AW excitation.

Interestingly, the sequence cracking/melting extended along the whole interface found in Experiment 1 and the intermixing of ice-water in Experiment 2 strongly differ from the de-icing phenomenology found for de-icing large aggregates of compact ice when excitation is induced with R-SAWs.[25] In that case, no extensive cracking was detected, and de-icing followed a lateral progression of a well-defined water-ice front until the completion of the melting process. A similar waterfront-ice interaction has also been claimed for SAW de-icing of highly porous rime ice formed by frosting.[30] The experimental findings in Figure 6 showcase that de-icing of large ice aggregates with Lamb waves occurs through a different mechanism, which can be summarized in the following two steps: i) ice cracking along a large zone of the ice-substrate interface; ii) progressive melting along the affected area, a process where ice and liquid water may appear mixed in the excited zones.

The simulations of the power injection of 120 µm R-SAWs into the ice have shown that this process is restricted to a small region of the interface. Once the water is formed in this zone, the progression of de-icing will depend on the efficiency of heat transmission through the lateral water/ice interface, a process whose efficiency would be limited by the small area of such interface available for heat diffusion and likely diminished by heat losses from the water zone into the air. Meanwhile, since the power coupling between the 510 µm Lamb wave and ice is less intense, the wave may affect a much larger ice-substrate interface. Since the transmitted mechanical energy is insufficient to induce a rapid melting, ice cracking has been found to be the first step of the de-icing process. Then, the heat caused by acoustothermal effects due to the electrical activation of the device would be transmitted through the entire cracked ice/substrate interface without losses into the air. Paradoxically, the less intense LN-ice AW mechanical coupling found for the 510 µm Lamb wave appears to render a more efficient de-icing process, at least for small ice aggregates, as indicated by the data in **Tables 2** and **3**. Further advances in this evaluation of results would require additional analysis of heat transmission through the different interfaces involved in the examined systems, a question that remains open for future investigations.

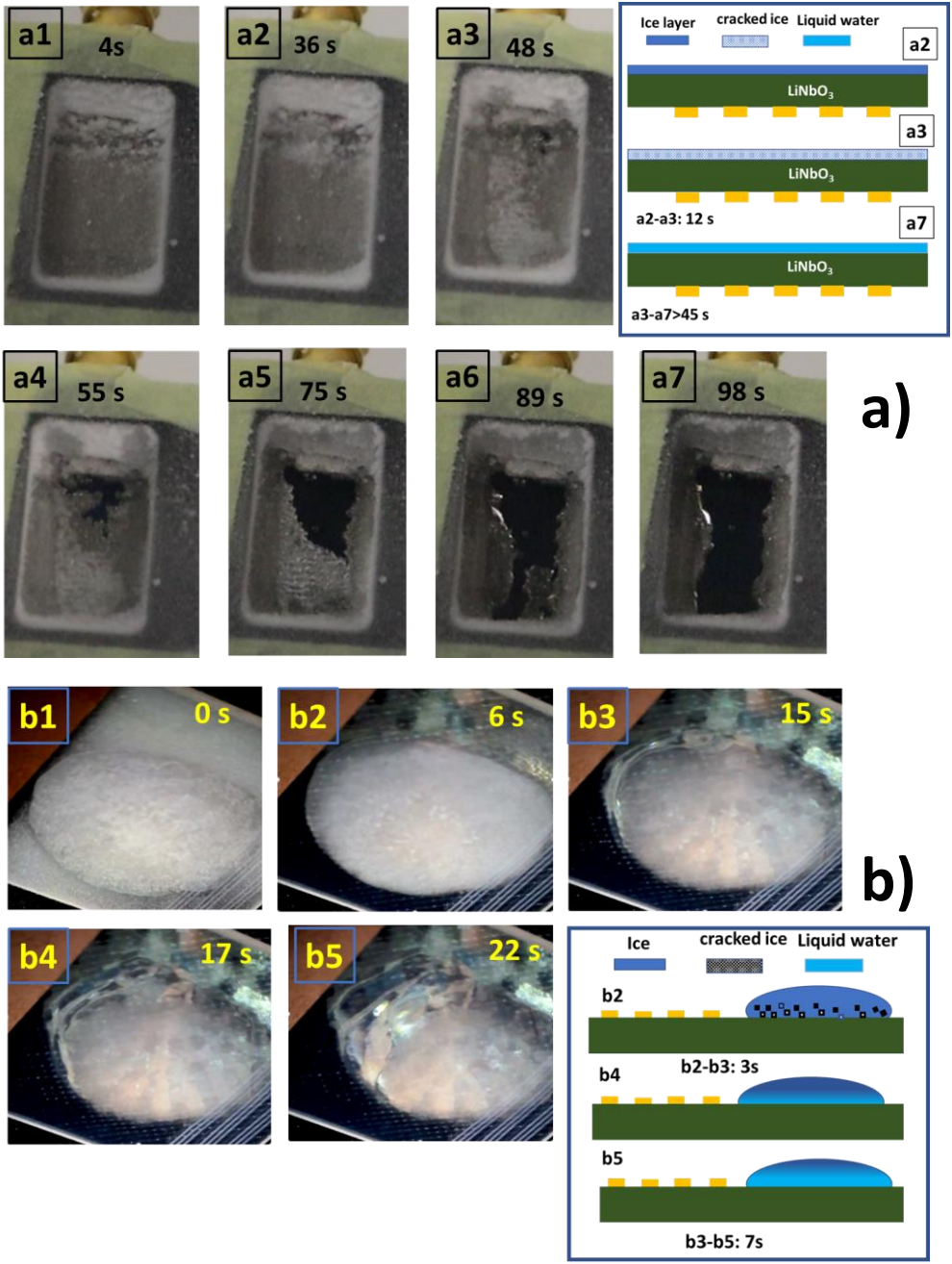

**Figure 6. Proof of concept experiments**: Experiment 1: a) Snapshot of a de-icing experiment induced with the 510 µm chip activated at 7.3 MHz with the IDTs at the back side of the LN chip as shown in Figures S5c and S5d; a1) Initial state with 1 mm of ice accreted on the surface of the chip; a2) Initial formation of cracks atop the IDT; a3) crack region extends to cover the whole chip in the direction of AW propagation; a4-a7) melting of ice on the zone atop the IDT. Inset: schematic of the de-icing mechanism induced by long wavelength Lamb waves.

Experiment 2: c) Snapshot of a de-icing experiment induced with the 510 µm chip activated at 7.3 MHz and a big ice aggregate facing the IDTs placed on the same side: b1) application of the AW; b2) ice change aspect without changes in the perimeter. Frost around disappears; b3) Progression of activation: some water forms at the edge facing the IDTs; b4 and b5) melting occurs in the whole aggregate, and water and ice coexist in the melted zones. Inset: schematic of the de-icing mechanism induced by long wavelength Lamb waves.

**CONCLUSIONS**

The experiments and simulations in this work have disclosed the different de-icing mechanisms affecting the interaction with ice of traveling AWs generated in piezoelectric substrates. It has also been determined that the de-icing efficiency, at least to melt small ice aggregates, is highly dependent on the characteristics of the AW. In particular, through the analysis and simulation of the interaction with ice of an R-SAW of relatively short wavelength and that of a long wavelength Lamb wave, significant differences have been determined in the phenomenology of de-icing for each type of wave. Thus, it has been shown that the former effectively transmits most of its mechanical energy to ice within a small area of the interface. At the same time, damping of the latter is less critical, and the Lamb wave may transit without a strong attenuation through a longer distance of the ice-substrate interface. Regarding the de-icing process, these specific properties of the AW-ice mechanical interactions lead, as a first step, to the direct melting of the ice edge when the activation is done with R-SAWs but to ice cracking within a large interface area when the activation is done with long wavelength Lamb waves. Progression of a water-ice front and interface melting over a large area are the successive steps of the de-icing process induced by these two types of AWs. Both types of AW and differences in wavelength seem to be essential features responsible for these differences. For the conditions explored in the present work, this combination of factors made de-icing and active anti-icing more favorable with Lamb waves. Additional studies should be carried out to determine the influence of each one of these characteristics quantitatively.

Reaching the previous conclusions has been possible by combining experiments with $LiNbO_3$ model systems and the finite element simulation of the AW-ice interaction. In this regard, results and simulations should be considered as a first step in investigating the use of AWs to prevent icing. Extrapolation of the findings reported here to other experimental conditions, types of waves, and icing conditions is not automatic, and specific experiments/simulations would be necessary to account for the phenomenology encountered in each case. In this regard, we deem that wavelength, type of wave (e.g., Lamb or Rayleigh, a different kind of surface waves or thickness shear waves), traveling or stationary waves, state of the surface, tilting angles of the substrates, etc., are critical characteristics that should be considered explicitly for a proper evaluation of de-icing mechanism and efficiencies.

**Supporting Information** ((delete if not applicable))

Supporting Information is available from the Wiley Online Library or from the author.

**Acknowledgments**

The authors thank the projects PID2022-143120OB-I00 and TED2021-130916B-I00 funded by MCIN/AEI/10.13039/501100011033 and by "ERDF (FEDER)" A way of making Europe, Fondos NextgenerationEU and Plan de Recuperación, Transformación y Resiliencia". CLS thanks the University of Seville through the VI PPIT-US and "Ramon y Cajal" program funded by MCIN/AEI/10.13039/501100011033. This research work is funded by the EU H2020 program under grant agreement 899352 (FETOPEN-01-2018-2019-2020 – SOUNDofICE).

**Table of contents**

Supporting information

**Mechanisms of de-icing by surface Rayleigh and plate Lamb acoustic waves**

*Shilpi Pandey,\* Jaime del Moral,\* Stefan Jacob, Laura Montes, Jorge Gil-Rostra, Alejandro Frechilla, Atefeh Karimzadeh, Victor J. Rico, Raul Kantar, Niklas Kandelin, Carmen López-Santos, Heli Koivuluoto, Luis Angurel, Andreas Winkler,\* Ana Borrás, Agustin R. González-Elipe.*

S. Pandey,*[+] S. Jacob, A. Karimzadeh, A. Winkler*

Leibniz IFW DresdenSAWLab Saxony,

RG Acoustic MicrosystemsHelmholtzstr. 20, 01069 Dresden, Germany

*E-mail: (shilpi.pandey@tum.de) (a.winkler@ifw-dresden.de)

S.Jacob

German National Metrology Institute (PTB), Bundesallee 100, 38106 Braunschweig, Germany

J. D. Moral,* L. Montes, J. G. Rostra, V. J. Rico, C. L. Santos, A. Borrás, A. R. González-Elipe.

Nanotechnology on Surfaces and Plasma Lab, Materials Science Institute of Seville, Consejo Superior de Investigaciones Científicas (CSIC), Americo Vespucio 49, Sevilla 41092, Spain

*E-mail: (jaime.delmoral@icmse.csic.es)

R. Kantar, N. Kandelin, H. Koivuluoto

Materials Science and Environmental Engineering, Faculty of Engineering and Natural Sciences, Tampere University, 589, 33104 Tampere, Finland

A. Frechilla, L. A. Angurel

Instituto de Nanociencia y Materiales de Aragón (INMA), CSIC-Universidad de Zaragoza, Zaragoza, Spain

*Current addresses*:

[+] Heinz-Nixdorf-Chair of Biomedical Electronics, School of Computation, Information and Technology. Technical University of Munich, TranslaTUM, 80333 Munich, Germany

**Supporting information S1**: **Mask design and IDT fabrication**

IDTs with wavelengths from 120 µm to 510 µm and a different number of finger pairs were fabricated as described in the main text of this article. The mask and IDTs distribution in Figure S1 exemplifies the optimization of the use of the wafer area to prepare a maximum number of chips.

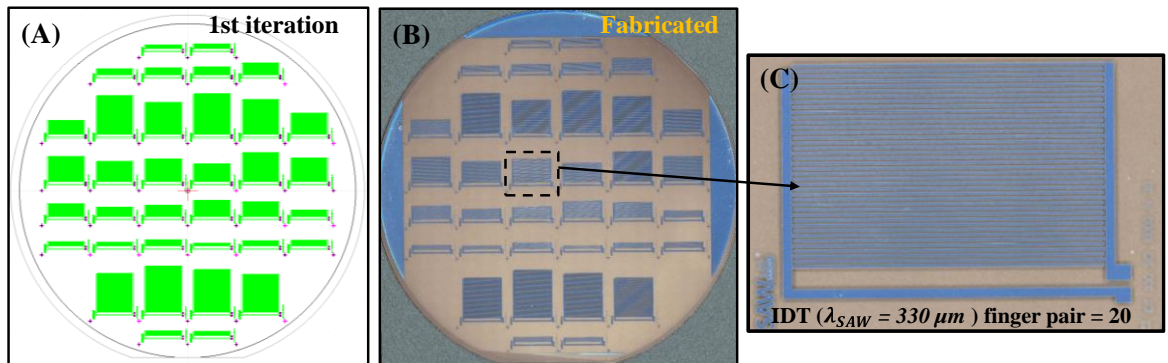

**Figure S1. Design and preparation of wafer layout for the AW chips. (A) Mask layout (CAD) for the fabrication of 1st iteration IDTs, (B) picture of fabricated wafer, (C) Micrograph of IDT for 330 μm wavelength and 20 finger pairs.**

**Supporting information S2: Wafer layout for dicing and AW chip holder assembly**

Selected chip devices for icing testing were specifically fabricated on a LN wafer and then diced. The chip devices were activated using a specially designed PCB board and holder.

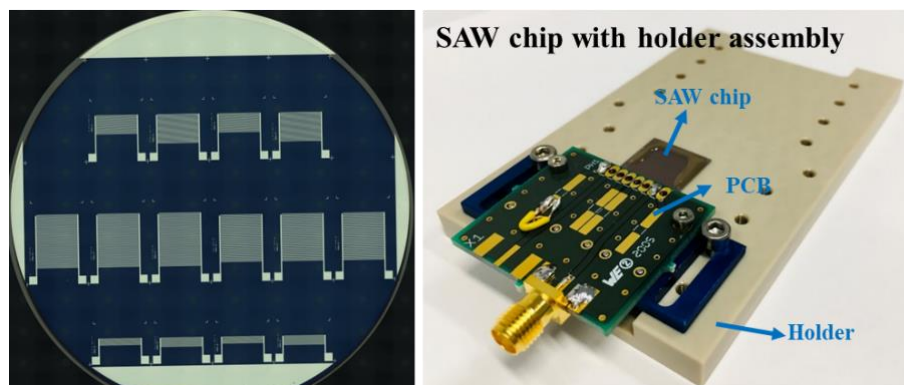

**Figure S2. Left) Fabricated SAW devices on a wafer black LiNbO$_3$ 128° YX cut single side polished (SSP) ready for dicing. Right) SAW chip emplaced on the devoted holder platform for de-icing and anti-icing experiments in the cool chamber.**

**Supporting information S3: Devices and procedures for testing the electroacoustic behavior of AW devices.**

SAW devices were electrically tested when still in the wafer by applying a resist to the wafer to electroacostically separate the chip. A suitable probe system was used for this testing.

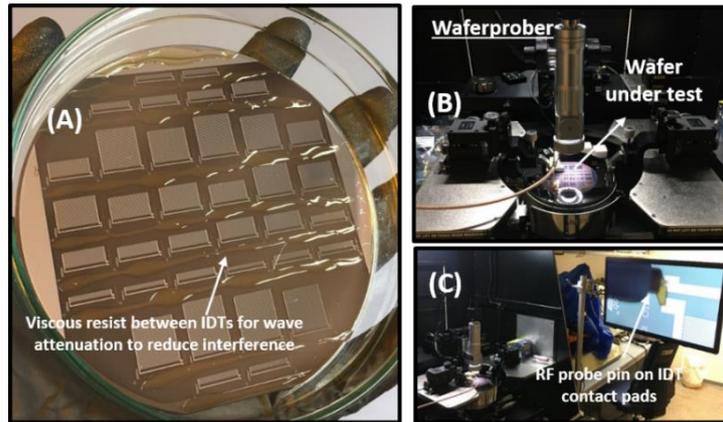

**Figure S3. Setup for electroacoustic characterization of the test chips. (A) Test wafer for electrical characterization with viscous resist added at the IDT edges to induce attenuation during RF measurements; (B & C): RF measurement setup, wafer under test.**

**Supporting Information S4: Ice chamber and water droplet system**

De-icing and active anti-icing experiments were carried out in an experimental stage consisting of a cooling chamber including a chip holder, a Peltier and a water dosing system. The electronic equipment is left outside the chamber and connected through suitable cables. Various thermocouples are distributed inside the chamber to monitor the temperature at critical points of the experimental stage.

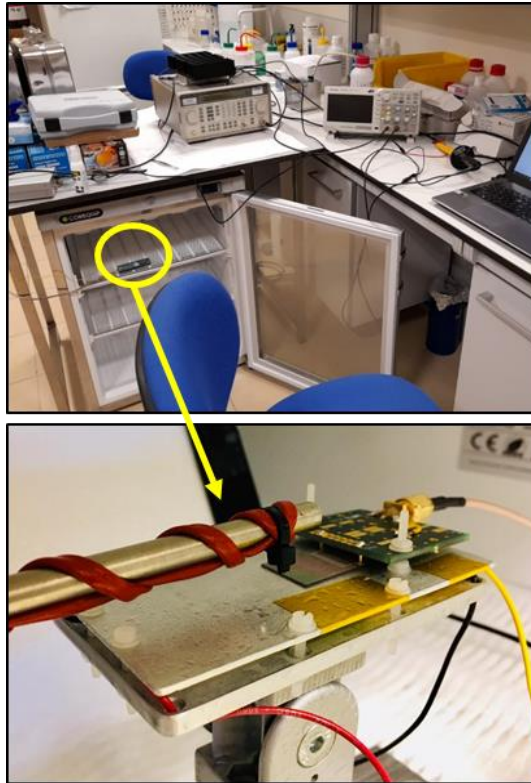

**Figure S4. Custom ice chamber and electrical setup for de-icing and anti-icing experiments.** Top) Photographs of the electronic equipment required for activation of AWs and the icing chamber (bellow the laboratory table) and experimental set up for holding and tilting the chip devices. Bottom) Detailed photograph showing the water hose, chip and tiltable Peltier plate holder.

**Supporting information S5: Housing and chip used for the de-icing experiment of a big layer of ice**

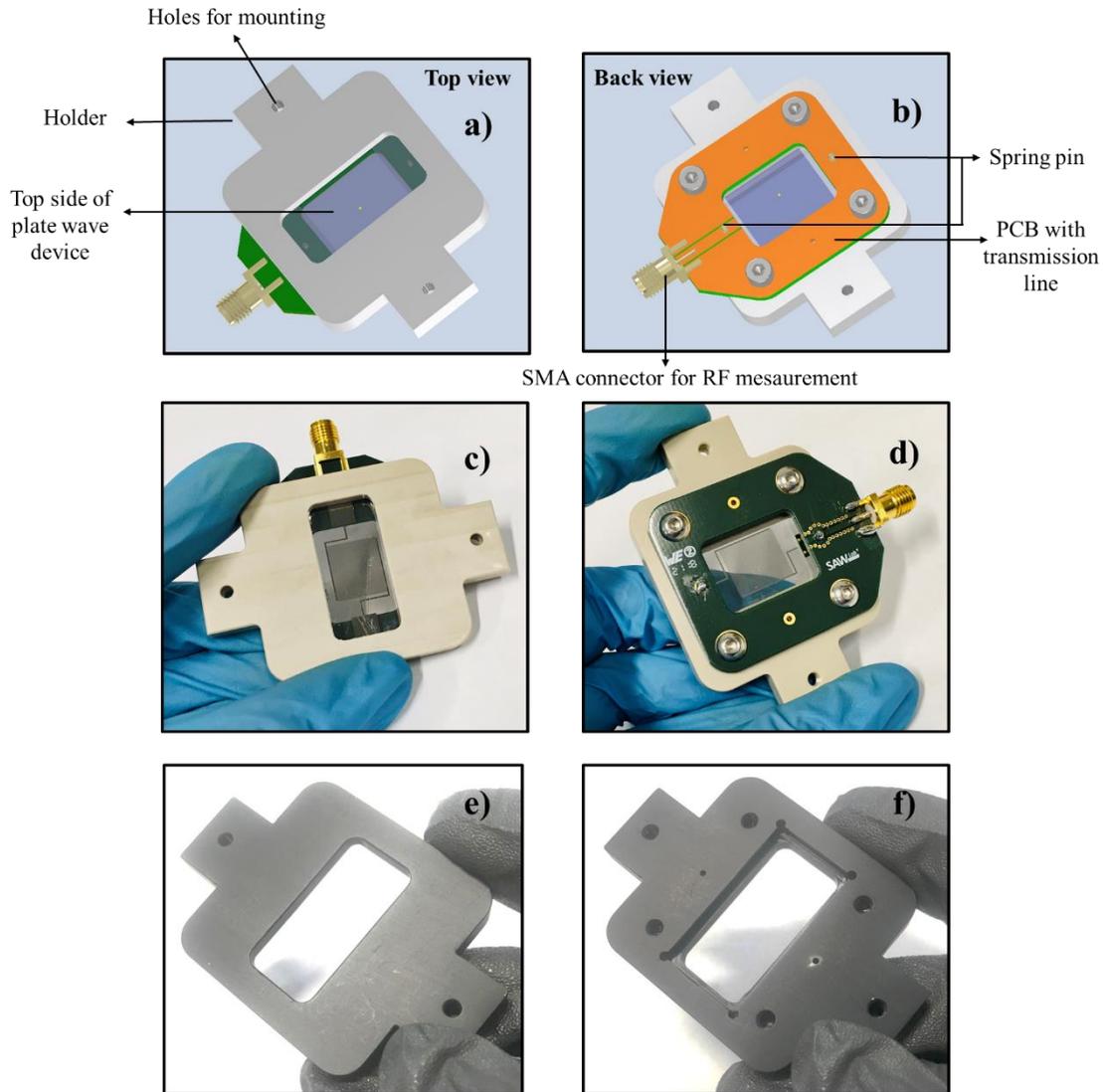

**Figure S5.** Housing and chip utilized for de-icing of a large layer of ice. a) and b) Schemes of holder and PCB used for the experiments. c) and d) Photographs of the chip incorporated in the holder and a description of the connections to the 510 µm. e) and f) Photographs of the protecting holder design to protect the chip and IDT from a direct exposure to ice/water.

Photographs in c) and d) show that the IDTs are placed in the center of the chip, on the opposite side to the face where ice accretion occurs during the experiments.

**Supporting information S6: Parameters used for the FE simulation of AW and the AW-ice interaction**

The values of the different parameters utilized for the FE simulations of the LN and ice ae included below:

### A) LiNbO₃ 128°Y-X physical properties and parameters

Density: $\rho = 4628\ Kgm^{-3}$
Elastic matrix (GPa) (Symmetric matrix: $C_{ij} = C_{ji}$)

$$C_{15} = C_{16} = C_{25} = C_{26} = C_{35} = C_{36} = 0$$
$$C_{11} = 198.4 \mid C_{12} = 66.307 \mid C_{13} = 53.4927 \mid C_{14} = 6.9567$$
$$C_{22} = 186.5753 \mid C_{23} = 80.4410 \mid C_{24} = 6.3254$$
$$C_{33} = 209.0427 \mid C_{34} = 6.0753$$
$$C_{44} = 75.0410 \mid C_{56} = 56.6210 \mid C_{66} = -3.9591$$

Piezoelectric matrix (Cm$^{-2}$)
$$e_{11} = e_{12} = e_{13} = e_{14} = e_{25} = e_{26} = e_{35} = e_{36} = 0$$
$$e_{21} = -1.7222 \mid e_{22} = 4.5332 \mid e_{23} = -1.3518 \mid e_{24} = 0.2173$$
$$e_{31} = 1.7263 \mid e_{32} = -2.4540 \mid e_{33} = 2.5952 \mid e_{34} = 0.7352$$

Relative permittivity ( Symmetric matrix: $\epsilon_{ij} = \epsilon_{ji}$)
$$\epsilon_{12} = \epsilon_{13} = 0$$
$$\epsilon_{11} = 45.5998 \mid \epsilon_{22} = 38.2846 \mid \epsilon_{33} = -9.3630$$
$$\epsilon_{23} = -9.3630 \mid \epsilon_{33} = 33.6157$$

### B) Isotropic 1h ice physical constants

Since for the ice aggregates domain, only the mechanical interactions have been simulated. No electromagnetic constants included in the model. Data are taken from ref. [39] in the main text.

Density: $\rho = 916 \: Kgm^{-3}$

Elastic matrix (GPa) (Symmetric matrix: $C_{ij} = C_{ji}$)
$$C_{11} = 6.82 \mid C_{12} = 5.35 \mid C_{22} = 6.82 \mid C_{33} = 6.82 \mid C_{44} = 3.41 \mid C_{55} = 3.41 \mid C_{66} = 3.41$$
The rest of the matrix elements are null.

### C) 2D FEM calculations

The 2D FEM simulations have been carried out in COMSOL 5.6. Characteristic parameters for calculations are shown in **Table S6**. These parameters are wavelength-dependent to optimize the results for each concerned AW. A direct consequence of this approach is that simulations for the 120 µm wavelength require a higher computing load than those for the 510 µm wavelength. This is due to the requirements of a finer mesh and a higher working frequency (shorter period and, therefore, a shorter time-step).

Characteristic parameters used for the simulations are summarized in **Table S6**. The perfect matcher layer (PML) thickness deserves an additional comment from them. PMLs are incorporated on the sides of the system geometry to avoid reflections. Inside them, a certain damping factor per length unit of length is applied on the lattice displacement field. If this PML domain is not thick enough, the lattice displacement will not be null at the ends of the chip, possibly causing undesired reflections. The thickness of PML should be carefully adjusted because a too-thick PML would mean a heavier computational load without further improvement in the results.

*Table S6.- Characteristic parameters used for the COMSOL simulations.*

| Name | Description | $\lambda = 120 \: um$ | $\lambda = 510 \: um$ |
|---|---|---|---|
| $N_\lambda$ | Number elements per wavelength | | 15 |
| $PML_{thick}$ | Thickness of the perfect matched layer | | $10\lambda$ |

| $t_{step}$ | Time-step for time dependent solver | 0.46 ns | 1.34 ns |
|---|---|---|---|
| $V_{in}$ | Voltage input on the IDTs for the time dependent simulations | 8.71 V | 10.00 V |
| $V_{ice}$ | Volume of the ice aggregate | 4 ul | |
| $f_{step}$ | Frequency step taken for the frequency domain simulations | 0.5 kHz | |

**Supporting information S7: Electrical characterization of AW devices with different wavelengths: Return loss spectra.**

Following the manufacturing methodology utilized to prepare the chips, devices operating at different wavelengths from 120 µm to 510 µm were manufactured and electrically characterized. De-icing experiments were carried out with the devices with the shortest and longest wavelengths. The first required a relatively smaller number of finger pairs to achieve an almost constant minimum value of the return losses (i.e., maximum use of the applied power to generate AWs). The $S_{11}$ curve of the 120 µm chips was the broadest one from the whole series, which made it easier for this device to tune the frequency for resonant conditions.

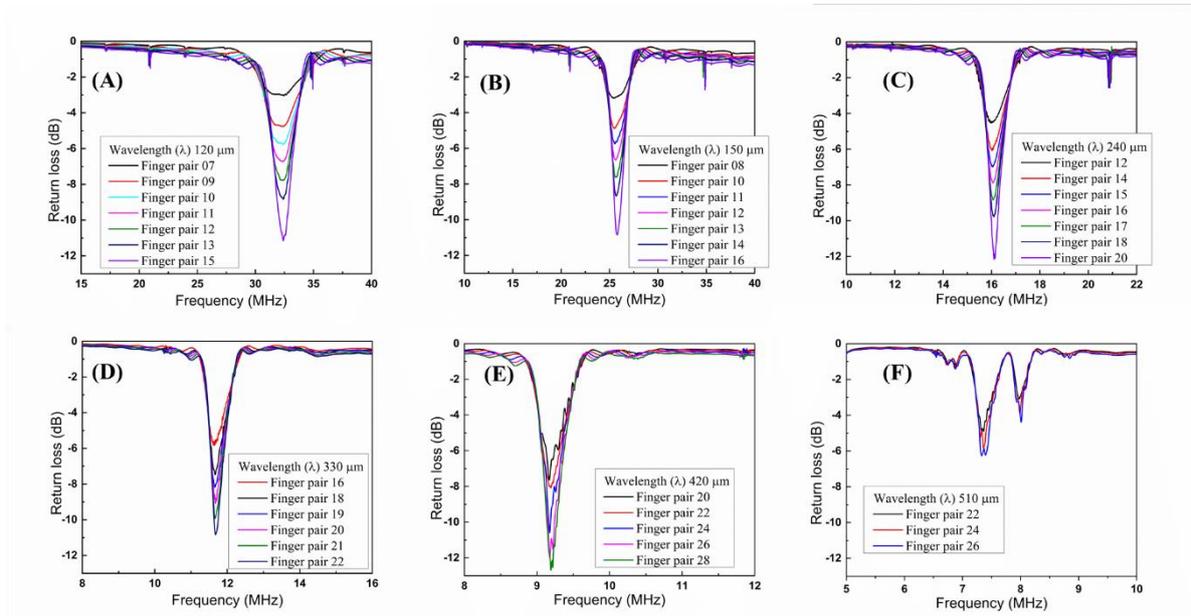

**Figure S7. Electrical characterization of SAW devices.** Return loss versus frequency plots for the fabricated AW devices for different wavelengths and number of finger pairs as indicated. (A) wavelength (λ) 120 µm; (B) wavelength (λ) 150 µm; (C) wavelength (λ) 240

µm; (D) wavelength (λ) 330 µm; (E) wavelength (λ) 420 µm; (F) wavelength (λ) 510 µm. Note teh different scales of x and y axis in each plot.

## Supporting information S8: Electrical characterization of AW devices with different wavelengths: reflection coefficient of power

The progressively lowering baseline found for most devices– ideally, $|S_{11}|^2$ equals 1 outside of the resonance peak – indicates that there exist electrical power losses on the order of 10-20%. These losses are expected to be caused by parasitic capacitances in the IDT and ohmic resistance of the thin film metallization, as well as other minor contributions e.g. from contact resistance. A higher number of finger pairs increases these losses, as expected. Secondly, the reflected power – defined by the $|S11|^2$ curve minima – could already be reduced to 5-10% throughout the wavelength range except for the longest wavelength, i.e. 510 µm.

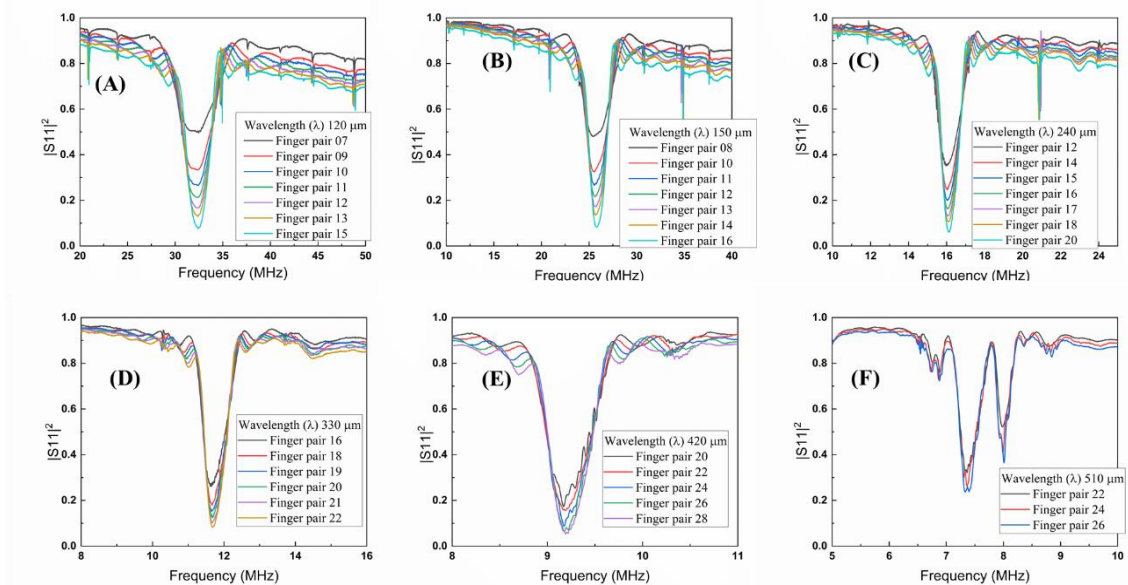

Figure S8. RF measurement results: Reflection coefficient of power $|S_{11}|^2$ versus frequency for devices with increasing number of finger pairs (indicated in the plots) and wavelengths. (A): wavelength (λ) 120 µm; (B): wavelength (λ) 150 µm; (C): wavelength (λ) 240 µm; (D) wavelength (λ) 330 µm; (E): wavelength (λ) 420 µm; (F): wavelength (λ) 510 µm.

**Supporting information S9: Generation of two wave Lamb modes for 600 um IDT chip devices.**

The generation of antisymmetrical and a symmetrical Lamb mode in LN chips activated with IDTs with wavelengths larger than the chip thickness is further demonstrated with this chip device activated with an IDT of 600 µm.

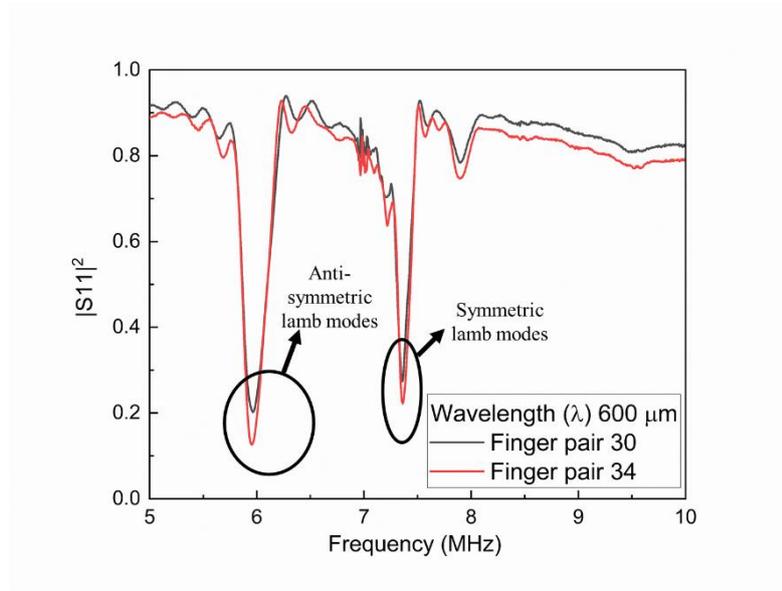

**Figure S9.** Plot of $S_{11}^2$ vs frequency for a chip device operated with 600 µm IDTs. Spectra for two set of finger pairs are included in the plot.

**Supporting information S10: Video showing the simulated 120 um R-SAW in the time domain.** Colour code and scale equivalent to that used for representing the snapshots in Figure 3 a2.

**Supporting information S11: Video showing the simulated A0 mode of the 510 um Lamb wave in the time domain.** Colour code and scale equivalent to that used for representing the snapshots in Figure 3 b2.

**Supporting information S12: Video showing the simulated S0 mode of the 510 um Lamb wave in the time domain.** Colour code and scale equivalent to that used for representing the snapshots in Figure 3 b2.

**Supporting information S13: Weak standing wave mode in the 510 um Lamb wave simulations**

An interesting aspect of the 510 µm simulation is that, in fact, the found wave mode presents a standing (STD) wave superpose on the S0 Lamb plate wave discussed in the main text. This phenomenon is depicted in Figure S12. Figure S12 a) shows the complete wave mode, where it is possible to find some areas of the wave where displacement is significantly higher. In Figure S12 c), a high-pass filter on the amplitude colormap has been applied to enhance this effect, meaning that all substrate points with a displacement modulus value under 0.9 nm use the same color. This filtering enables a better color-resolution for points with high enough amplitudes, allowing us to see clearly the STD wave. This wave shows a wavelength of 4 mm. The origin of this simulated contribution is controversial and might be due to back-and-forth reflections/interferences of the main contribution on the chip borders. Since the magnitude of this possible modulation is not high, we do not consider it of significance from an experimental point of view.

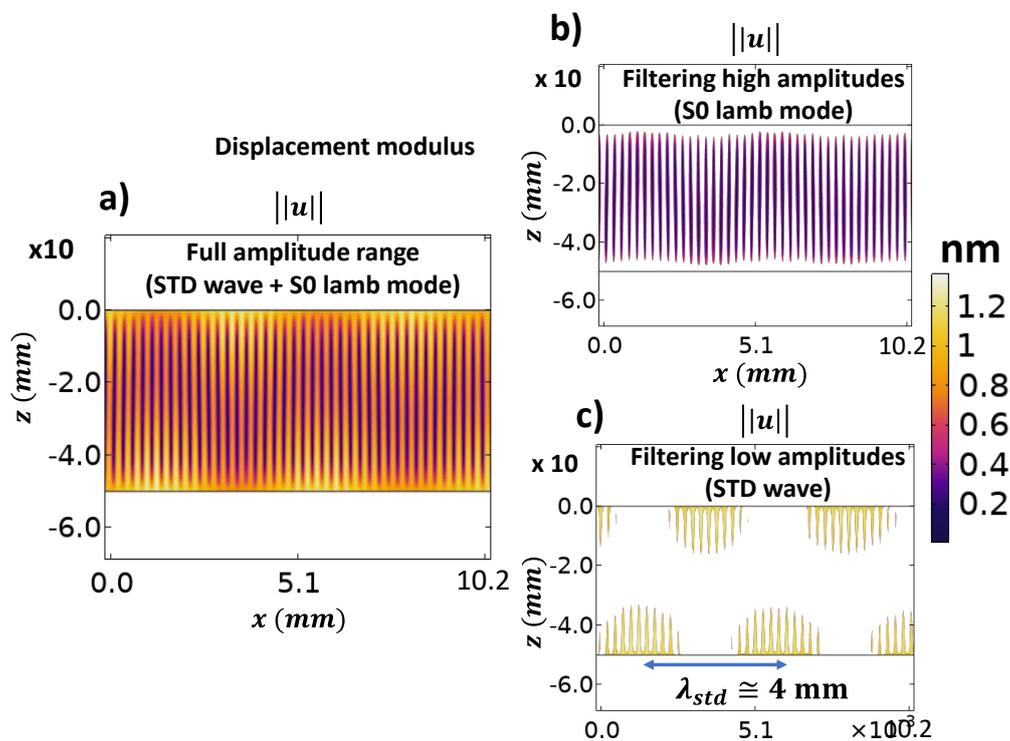

**Figure S13. Analysis of the standing wave present in the 510 μm wavelength simulations; a) full displacement modulus colormap; b) displacement colormap with a low-pass filter on the displacement values, therefore only displacement values under the threshold are represented; c) displacement colormap with a high-pass filter on the displacement values, only values over the threshold are represented.**

**Supporting information S14: De-icing of small aggregates at -5ºC.**
Snapshots of these experiments are included here as complement of similar snapshots taken for the de-icing process at -15ºC (c.f. Figure 4 in the main text).

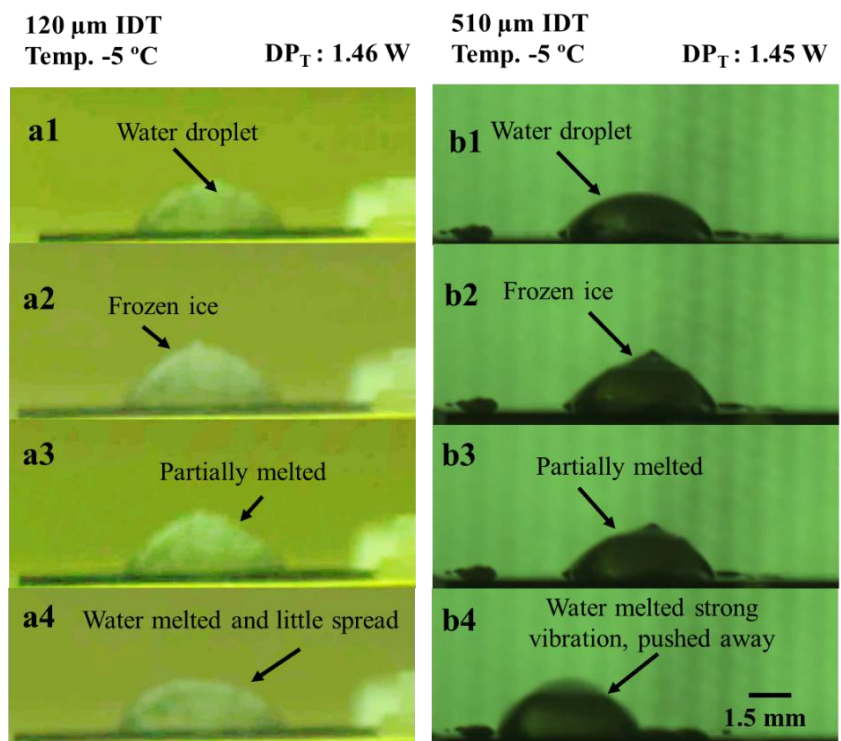

**Figure S14.** Comparison of de-icing process on the 120 µm (a) and 510 µm (b) devices at -5ºC: a 22 µL droplet is delivered on the chips (a1, b1); droplet is frozen; (a2, b2): droplet starts melting, after chip activation with the DPt value reported in table 2 (a3, b3); droplet remains liquid (active anti-icing effect) (a4,b4). The behavior of the melted water droplet was different for the two devices: it spread for the 120 µm device but was pushed toward the chip edge in the case of the 510 µm device. The water droplet was dispensed at almost a similar distance from the last finger of the IDTs (i.e., at about 3 mm).

**Supporting information S15. Video showing the melting of the water droplet at -15 ºC using the 120 µm R-SAW**

**Supporting information S16. Video showing the melting of the water droplet at -15 ºC using the 510 µm Lamb wave R-SAW**

**Supporting information S17. Video showing the simulations in the time domain of the interaction of the 120 µm R-SAW with an ice aggregate.** Colour code and scale equivalent to that used for representing the snapshots in Figure 5 a.

**Supporting information S18. Video showing the simulations in the time domain of the interaction of the 510 µm Lamb wave with an ice aggregate.** Color code and scale equivalent to that used for representing the snapshots in Figure 5 b.

**Supporting information S19.- Simulations of the interaction with ice of 510 µm R-SAW and 510 µm Lamb waves**

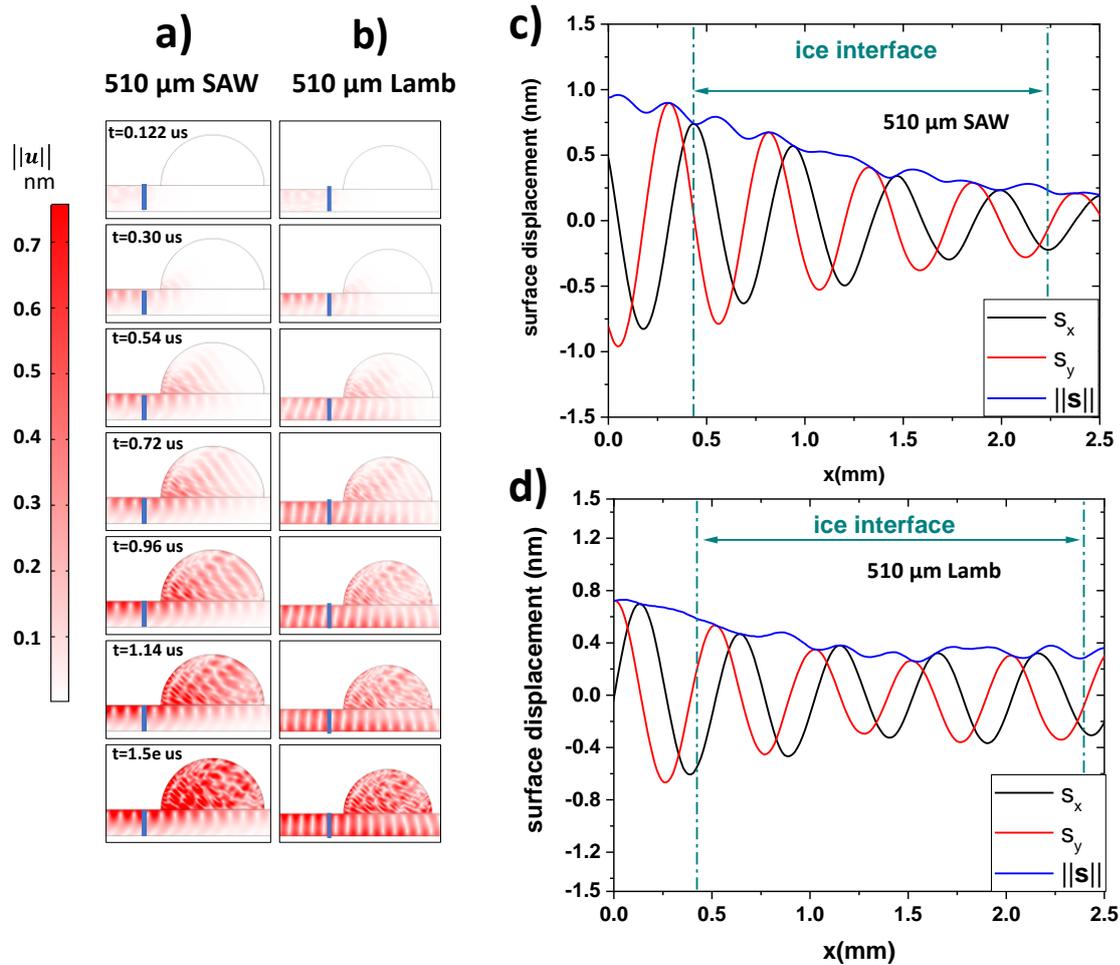

**Figure S19. Comparison of AW-ice interaction and surface displacements for 510 µm R-SAW and 510 µm Lamb waves.** Time-dependent simulated snapshots in the form of intensity color maps of the AW-induced displacements in the LN chips with an ice aggregate on their surface. The vertical line along the plate signal the location of the last finger of the simulated IDT. a) and b) Calculations for the 510 µm R-SAW and 510 µm Lamb wave, correspondently. c-d) Surface displacements at the LN substrate as a function of wave propagation along the x coordinate direction for the 510 µm R-SAW (c) and 510 µm Lamb wave (d) interacting with the ice aggregate on the surface.

**Supporting information S20. Video showing the simulations in the time domain of the interaction of the 510 µm R-SAW with an ice aggregate.** Colour code and scale equivalent to that used for representing the snapshots in Figure S19.

**Supporting information S21. Video of de-icing of a large ice layer covering a device activated with 510 µm Lamb waves**. Experiment 1 in main text.

**Supporting information S22. Video of de-icing of a large ice aggregate covering a device activated with 510 µm Lamb waves** Experiment 2 in the main text

**VIDEOS UPON REQUEST TO THE AUTHORS**